\documentclass[12pt]{aastex63}
\usepackage{graphicx}
\usepackage{amssymb}
\usepackage{epstopdf}
\usepackage{natbib}
\usepackage{subfigure}
\usepackage{epsfig}
\usepackage{multirow}
\usepackage[figuresright]{rotating}
\usepackage{amsmath}

\begin{document}

\title{Habitability and Water Loss Limits on Eccentric Planets Orbiting Main-Sequence Stars}
\correspondingauthor{Igor Palubski}
\email{ipalubsk@uci.edu}

\author{Igor Z. Palubski}
\affiliation{Department of Physics and Astronomy, University of California, Irvine, 4129 Frederick Reines Hall, Irvine, CA 92697}

\author{Aomawa L. Shields}
\affiliation{Department of Physics and Astronomy, University of California, Irvine, 4129 Frederick Reines Hall, Irvine, CA 92697}

\author{Russell Deitrick}
\affiliation{Center for Space and Habitability, University of Bern, Gesellschaftsstrasse 6, 3012 Bern, Switzerland}

\begin{abstract}
A planet's climate can be strongly affected by its orbital eccentricity and obliquity. Here we use a 1-dimensional energy balance model modified to include a simple runaway greenhouse (RGH) parameterization to explore the effects of these two parameters on the climate of Earth-like aqua planets \textemdash completely ocean-covered planets\textemdash orbiting F-, G-, K-, and M-dwarf stars. We find that the range of instellations for which planets exhibit habitable surface conditions throughout an orbit decreases with increasing eccentricity. However, the appearance of temporarily habitable conditions during an orbit creates an eccentric habitable zone (EHZ) that is sensitive to orbital eccentricity and obliquity, planetary latitude, and host star spectral type. We find that the fraction of a planet's orbit over which it exhibits habitable surface conditions is larger on eccentric planets orbiting M-dwarf stars, due to the lower broadband planetary albedos of these planets. Planets with larger obliquities have smaller EHZs, but exhibit warmer climates if they do not enter a snowball state during their orbits. We also find no transient runaway greenhouse state on planets at all eccentricities. Rather, planets spend their entire orbits either in a RGH or not. For G-dwarf planets receiving 100\% of the modern solar constant and with eccentricities above 0.55, an entire Earth ocean inventory can be lost in 3.6 Gyr. M-dwarf planets, due to their larger incident XUV flux, can become desiccated in only 690 Myr with eccentricities above 0.38. This work has important implications for eccentric planets that may exhibit surface habitability despite technically departing from the traditional habitable zone as they orbit their host stars.
\end{abstract}

\section{Introduction}
With the rapidly expanding catalog of discovered exoplanets, much effort will be dedicated to characterizing these planets and identifying those that may be habitable\textemdash that is, possessing conditions conducive to the presence of liquid water \citep{Kasting1993, Kopparapu2013c, Seager2013}. Habitability is strongly dependent on many stellar, orbital, and planetary parameters \citep{Meadows2018b, Shields2019}. A first-order approach to identifying a potentially-habitable planet is to pinpoint one that orbits within the boundaries of its host star's habitable zone\textemdash the region around a star where a planet with an Earth-like atmosphere may be warm enough for liquid water to flow on its surface \citep{Kasting1993}. The inner edge of the habitable zone (IHZ) is determined by the onset of the runaway greenhouse, a climate state in which the atmosphere becomes opaque to outgoing thermal (longwave) radiation, inhibiting a planet's ability to cool and desiccating the surface, leaving zero water content on the planet. At the outer edge of the habitable zone (OHZ), determined by the maximum CO$_2$ greenhouse (GH) limit, any further addition of CO$_2$ into the atmosphere will no longer keep surface temperatures above the freezing point of water. \citep{Kasting1993, Pierrehumbert2010, Kopparapu2013a, Kopparapu2013b}. However, the traditional boundaries of the habitable zone are based on the assumption of Earth-like planetary conditions and do not take into account the range of orbital eccentricities or obliquities possible in extrasolar planetary systems. The large variations in orbital distance from their stars of highly eccentric planets may generate significant changes in surface temperature, creating intervals of habitable surface conditions interspersed with climate extremes during an orbit, defying traditional habitable-zone calculations \citep{Linsenmeier2015}. Similarly, large planetary obliquities lead to larger seasonal variations in the stellar flux latitudinal pattern, which in turn can drastically affect a planet's climate and possibly push it permanently into a snowball state. Counter to the climatic state of present Earth, at an obliquity of 23.44$^\circ$, if a planet's obliquity is 54$^\circ$ or greater  (\citealp{Williams1975}), the polar regions receive more stellar flux and tend to be warmer than the equatorial regions, leading to the formation of ice belts\textemdash ice covered regions that extend from the equator poleward. Below this threshold ice caps\textemdash ice-covered polar areas of a planet\textemdash are formed instead. In general, increasing obliquity destabilises the ice caps, i.e., at higher obliquities, ice caps collapse to the equator at higher values of stellar flux \citep{Rose2017}. Similarly, on planets with obliquities higher than 54$^\circ$, the corresponding ice belts collapse towards the poles. A planet's temporal habitability\textemdash defined here as any fraction ($<$ 1) of the orbital period over which habitable conditions are present\textemdash as a function of its orbital eccentricity has not previously been quantified. Temporarily habitable planets may experience a snowball or a runaway greenhouse (RGH) state over a significant portion of the orbit but are habitable for its remainder. Surface life on such planets would likely have to seek shelter through a RGH period and/or hibernate through a snowball period. The survival of subsurface life through a snowball episode will depend on the thickness of sea ice. If ice grows to a few hundred meters or more, photosynthesis will not be possible. However, if it remains thin, or if there exist some oases\textemdash small deglaciated regions\textemdash life may survive these snowball episodes \citep{Abbot2013}.

Many confirmed exoplanets are on eccentric orbits (see, e.g., \citealp{Korzennik2000, Naef2001, Jones2006, Giguere2012, Harakawa2015, Kane2016, Wittenmyer2017, Brady2018}). At the time of writing, $\sim$50\% of these have orbital eccentricities $e >$ 0.1 and $\sim$10\% have $e >$ 0.5.\footnote{The extrasolar planets encyclopedia. 〈http://www.exoplanet.eu/} Previous studies on the effects of eccentricity on the habitability of an Earth-like planet orbiting a Sun-like star found that such planets may have liquid water on the surface even at high eccentricities \citep{Williams2002, Dressing2010}. These studies determined that the habitability of an eccentric planet may be approximated by the annually-averaged stellar flux received by the planet, which may correspond to a distance that is within the boundaries of the traditional habitable zone \citep{Williams2002}. High eccentricity may therefore help planets maintain habitable surface conditions near or even outside of the OHZ \citep{Kopparapu2013a, Kopparapu2013b}, though the highest eccentricities have been shown to induce cyclic snowball climate behavior \citep{Bolmont2016a}. And smaller gaseous, ``mini Neptune" exoplanets on eccentric orbits may undergo photoevaporation of their hydrogen/helium envelopes, revealing potentially habitable Earth-mass planets \citep{Luger2015a}. Similarly, obliquity has been shown to have a significant impact on planetary climate. Large obliquity may pose difficulties for habitable climates due to ice instabilities, but high-obliquity habitable Earth-like planets are possible \citep{Williams1997b, Spiegel2009, Armstrong2014, Ferreira2014}. These studies underscore the importance of quantifying the effects of extreme orbits on planetary habitability.

The climatic effect of eccentricity has been studied using a one-dimensional (1-D) energy balance model (EBM) \citep{Dressing2010}. Planets have been found to remain habitable for a range of eccentricities, and when initially frozen, thaw if perturbed to a higher eccentricity \citep{Dressing2010}. However, this work did not include a parameterization to simulate the runaway greenhouse state to which highly eccentric planets are susceptible at periastron (closest approach to the star), nor did it quantify temporal habitability as a function of the spectral energy distribution (SED) of a planet's host star. Planets in a runaway greenhouse state have surface temperatures exceeding the critical point of water (647 K), leading to complete evaporation of their oceans. Since there is no cold trap at such extreme temperatures, all water vapor rises unrestrictedly into the upper layers of the atmosphere, where it can be photolyzed. While hydrogen more easily escapes to space, oxygen may remain behind to form ozone, to oxidize the surface, or to build up O$_2$-rich atmospheres that may present a false positive signature for life \citep{Wordsworth2014, Luger2015b}.

An existing complication to the potential habitability of M-dwarf planets remains the extreme activity of their host stars \citep{Scalo2007, Tarter2007, Shields2016b}. M-dwarfs emit strongly in the X-ray (0.1-12 nm) and extreme UV (12-1000 nm) regions of the electromagnetic spectrum (hereafter ``XUV"). XUV photons can drive the gravitational escape of atmospheric constituents \citep{Luger2015b, Lammer2007, Erkaev2007, Lammer2009}. Stellar winds, coronal mass ejections, and flare activity can exacerbate these effects \citep{Lammer2007, Khodachenko2007, Odert2017}. Planets with sufficiently large water mixing ratios in the atmosphere are particularly susceptible to desiccation of their surfaces given host stars with high amounts of XUV flux.

For this study we use a 1-D, latitudinally resolved EBM with an explicit sea ice model and a RGH parameterization to investigate the effects of eccentricity and obliquity on the climate and habitability of terrestrial aqua planets \textemdash completely ocean-covered planets \textemdash orbiting F-, G-, K-, and M-dwarf stars. The eccentric habitable zone (EHZ) considers the orbit-averaged flux as the main predictor of habitability on eccentric planets \citep{Barnes2008}. The EHZ compares the orbit-averaged flux on an eccentric orbit to the flux values within the classical habitable zone, which corresponds to a circular orbit. Here, we refine the definition of the EHZ to include the effects of strong seasonality. Previous work found that planets orbiting cool, lower-mass, M-dwarf stars, whose SEDs peak at longer, redder wavelengths, are more stable against global glaciation, and thaw out of such states at lower levels of incoming stellar radiation (hereafter ``instellation") compared with planets orbiting hotter, more luminous stars with more visible and near-UV output \citep{Shields2013, Shields2014}. The effect of host star SED on the climate and habitability of eccentric planets may therefore be significant, and has not yet been constrained. While Barnes \emph{et al.} (\citeyear{Barnes2008}) calculated the EHZ as a function of effective stellar temperature, an exploration of the climatic effect of host star spectral energy distribution (SED) was not included.

We account for periods of both snowball and moist/runaway greenhouse and calculate the fraction of the planetary surface that has clement conditions for liquid water throughout its orbit. We do this for planets orbiting stars of different spectral type, assuming a fixed (Earth-like) amount of atmospheric CO$_2$. We also calculate the full-orbit EHZ, which corresponds to planets that exhibit habitable conditions over the entire orbit. Lastly, we calculate the water loss rates for planets in an eccentricity-instellation parameter space where they are subject to both runaway and moist greenhouse states. We compare the timescales for these planets to lose an entire Earth ocean inventory as a function of their host star spectral type.

In Section 2 we describe the modifications made to the EBM to implement the RGH parameterization, as well as the model we use to calculate the water loss rates for different stellar XUV fluxes. In Section 3 we present the results in the form of EHZ instellation ranges, habitability fractions, water loss rates, and ocean loss timescales, as a function of eccentricity and host star SED. In Section 4 we discuss the implications of this work for the habitability of planets whose orbits take them interior to and well outside of the traditional boundaries of the habitable zone. Conclusions follow in Section 5.

\section{Methods}

We use a 1-D Energy Balance Model (EBM), based on North and Coakley (\citeyear{North1979}), that has been used to explore the potential climates of exoplanets in previous work \citep{Shields2013}. This seasonally-varying model balances the absorbed incident stellar energy flux with the outgoing longwave flux and horizontal heat diffusion at all latitudes. As a 1-D latitudinal model that averages over longitudes, the EBM inherently applies best to rapidly rotating planets. Here we assume that our modeled planets are rapidly rotating like the Earth, where the rotational frequency is much larger than the orbital frequency. The original model was modified to include a latitudinally-varying diffusion coefficient that adjusts tropical heat transport to generate temperatures consistent with thermal wind observations \citep{Lindzen1977}. Our EBM includes an explicit sea ice model, where the ocean is allowed to freeze once temperatures drop below -2$^\circ$ C, producing either ice caps or ice belts, depending on planetary obliquity. The model incorporates the energy flux between the ocean and ice but no ice dynamics.

The outgoing longwave radiation (OLR) is linearly parameterized with surface temperature based on the effects of CO$_2$ and water vapor on the radiative properties of the atmosphere. An atmosphere with a condensable greenhouse gas has been shown to have a linear scaling of the OLR \citep{Koll2018}. This linear scaling is relatively independent of the water content for as long as some water is present, but flattens with increasing temperatures due to the disappearance of the spectral window regions. We modify the EBM to include a parameterization for a RGH limit as follows: At a surface temperature of 46.3$^\circ$ C and a corresponding OLR of 300 W/m$^2$ (the Komabayashi-Ingersoll Limit \citep{Ingersoll1969}), we hold the OLR fixed as surface temperatures continue to increase, to simulate the atmosphere's opacity to IR radiation, characteristic of a RGH effect \citep{Ingersoll1969}:

\begin{equation}
    OLR = 
         \begin{cases}
          A + BT & T \leq 46.3 ^{\circ} C\\
          300 \, W/m^2 & T > 46.3^{\circ} C, \\
         \end{cases}
\end{equation}
where $A = 203.3 \frac{W}{m^2}$ and $B = 2.09 \, \frac{W}{m^2 \, ^{\circ}C}$. We run model simulations until the annual rate of change in global mean surface temperature falls below $0.004^{\circ}\frac{C}{yr}$, at which point the model is designated as converged. However, planets that enter the RGH do not converge, and we end these simulations once the mean global temperature exceeds $100^\circ C$. This approach allows each model run to reach equilibrium while maintaining computational efficiency.
We simulate aqua planets by assigning a uniform distribution of 99\% ocean and the smallest (1\%) percentage of land required to prevent singular behavior in the model. For all model simulations we assume a rapid (24-hr) planetary rotation rate, an orbital period of 360 days, and an incident flux of 1360 W/m$^2$ at the averaged Earth-Sun distance at zero eccentricity, to isolate the effects of orbital eccentricity and host star SED on planetary climate. Planets orbiting in the habitable zones of lower-mass stars may be captured into 1:1 spin-orbit resonances \citep{Dole1964, Kasting1993, Joshi1997, Edson2011, Shields2016b}, which will certainly affect climate (see, e.g., \citealp{Showman2013}). However, highly eccentric planets, which are the focus of this study, are more likely to exhibit a higher order spin-orbit resonance than synchronous rotation \citep{Dobrovolskis2007}.

\subsection{Water Loss and Runaway Greenhouse}

We calculate the water loss rates for planets with Earth-like atmospheres and G- or M-dwarf host stars, via the energy-limited escape mechanism \citep{Selsis2007, Luger2015b, Heller2015, Bolmont2017}, which allows us to place the strongest constraints on ocean inventory loss rates. We characterize planets in a moist greenhouse as those having atmospheric water mixing ratios between $3\times10^{-3}$ and 1, the upper limit being the point when a RGH ensues \citep{Kasting2015, Wolf2015, Wolf2017a}. Moist greenhouse planets have stratospheric temperatures high enough to raise the cold trap higher up in the stratosphere or remove it completely. We estimate the water loss rate of a planet due to its host star's XUV flux as a function of orbital eccentricity.

We use a similar prescription to that of prior work \citep{Selsis2007, Bolmont2017}, where planetary water loss varies with host star XUV flux at a given orbital distance. We expand this framework to any orbital distance on an eccentric orbit. The following changes are made to estimate the order of magnitude of water loss for moist and RGH planets: First, we identify three temperature regimes which correspond to different water vapor mixing ratios based on the work of Kasting \emph{et al.} (\citeyear{Kasting2015}) and Wolf \emph{et al.} (\citeyear{Wolf2015}). For temperatures \textit{$T < 340$}, the cold trap appears within the stratosphere, preventing any significant mass loss \citep{Kasting2015}. For the $340 < T < 350$ bracket, we adopt a water mixing ratio of $3\times 10^{-3}$, and in the  $350 < T < 370$ bracket a ratio of $10^{-1}$. Finally, for $T > 370K$ the water mixing ratios approach unity \citep{Kasting2015}. Figure 1 shows an example evolution of the stratospheric water content and the mass loss rate of a model G-dwarf planet with e = 0 and $S = 125\% S_0$. This particular planet enters the RGH after 35 years, and it goes through all four brackets of stratospheric water content, starting from a dry stratosphere up to water mixing ratios approaching unity. The uptick in the mass loss rate at the end of the simulation is the actual mass loss rate in a RGH, but we end the simulations before water mixing ratios approach unity as the final climatic state is known. The first and second brackets correspond to a low and high mixing ratio moist greenhouse. The fourth temperature bracket denotes water loss in planetary regions where the mixing ratios approach unity while the global climate remains stable. This approach allows us to pinpoint those planets in our simulations whose surface conditions were likely indicative of moist greenhouse atmospheres. All temperature regimes and corresponding mixing ratios used in our model are listed in Table 1.

We identify RGH planets as those with surface temperatures exceeding 100$^\circ$ C (such that the water vapor mixing ratio is $\sim$ 1), and no equilibrium state within 250 model years of simulation. The mass loss rate of the atmosphere $\dot{m}$ is calculated by equating the absorbed energy of XUV photons to the gravitational potential at the surface of the planet. Similar to Scalo \emph{et. al} (\citeyear{Scalo2007}) and Bolmont \emph{et. al} (\citeyear{Bolmont2017}), we link the XUV flux at 1 AU to the mass loss rate of the atmosphere at any distance. However, for each temperature regime of the moist greenhouse, we add a multiplicative factor proportional to the water vapor mixing ratio in the upper atmosphere, and multiply by the corresponding photon absorbing area. Finally, we sum up the mass loss contribution from each region and integrate over the course of the orbit and divide by the orbital period to get the annual mass loss of the atmosphere:

\begin{equation}
    \dot{m} = \frac{1}{P} \sum_{n=1}^{3} \int_{P} \frac{\epsilon\, \kappa(t,\lambda)  \, F_{XUV}[d(t),\lambda]
     \cdot A_n[t, \lambda] R_p}{GM_p} \, dt
\end{equation}

where $M_p$ and $R_p$ are the planet's mass and radius, $d(t)$ is the star-planet distance at time t, $\lambda$ is the latitude, P is the orbital period, $\epsilon$ is the XUV absorption efficiency\textemdash the fraction of incoming XUV energy transformed into gravitational potential through mass loss, $\kappa$ is the water mixing ratio factor and $A_n$ is the surface area of each water mixing ratio regime. For all our simulations, $R_p = R_\oplus$, $M_p = M_\oplus$.
Planets in a RGH are located within the highest temperature regimes of the sample, with mixing ratios of 1 across the entire planet, reducing the above expression to just one term:

\begin{equation}
\dot{m} = \frac{1}{P} \int_{P} \frac{\epsilon F_{XUV}[d(t),\lambda] \pi R_p^3}{GM_p} \, dt
\end{equation}

The ocean loss rate is 9 times larger than the hydrogen escape rate, due to the stoichiometry of the photo-dissociation of water (i.e., for every two hydrogen atoms escaping the atmosphere, a water molecule, which weighs $\sim$9x as much, must be photo-dissociated). Following the derivation in Luger and Barnes (\citeyear{Luger2015b}), the critical flux at which oxygen begins to escape is:

\begin{equation}
F_{crit} = 180 \ \left(\frac{M_p}{M_\oplus} \right)^{2} \ \left(\frac{R_p}{R_\oplus} \right)^{-3} \ \left(\frac{\epsilon}{0.3}\right)^{-1} erg \ cm^{-2} \ s^{-1} ,
\end{equation}

which corresponds to $F_{crit} =$ 0.54 W/m$^2$ for Earth-like planets and an absorption efficiency factor of 0.1. Given a semi-major axis of 1AU, such high XUV flux values are only attained at periastron passage at eccentricities above $e = 0.767$ and $e = 0.907$ in the M- and G-dwarf planet cases, respectively. As these eccentricities constitute a small fraction of the total parameter space explored, we assume here that only hydrogen escapes the planet's gravitational well while the oxygen remains behind.

\subsection{Fractional Habitability}
Modeling efforts typically base surface habitability on a planet's annually-averaged global surface temperature. By this metric, habitability can be ``lost" over the course of the orbit of an eccentric planet, particularly during the farthest (apoastron) and closest (periastron) approaches to the host star. As the planet moves farther out towards apoastron, it can completely freeze over. Conversely, at periastron, surface temperatures can reach high enough levels for the planet to enter a moist or RGH state. Between these orbital extremes, an eccentric planet may exhibit temporal habitability, with clement conditions for surface liquid water at times during its year. To quantify the amount of temporal habitability on eccentric planets, we adopt the ``fractional habitability" approach of Spiegel \emph{et al.} (\citeyear{Spiegel2008}), where the ``habitability function", $H[d(t),\lambda]$, is equal to one for latitudes with habitable temperatures at a given position in the orbit, and zero otherwise:

\begin{equation}
    H[d(t),\lambda] = 
     \begin{cases} 
      1 & 270 \, \leq \, T(\lambda,t) \, \leq \, 370\ K\\
      0 & \text{all other temperatures} \\
   \end{cases}
\end{equation}

The fraction of the year for which each latitude is in the habitable temperature range, $f_{time}$ (the latitudinal fractional habitability), is the time-integrated habitability function divided by the orbital period:

\begin{equation}
    f_{time}[\lambda] = \int_P \frac{H[d(t),\lambda]}{P} dt
\end{equation}

Finally, the net fractional habitability is the area-weighted integral of the latitudinal fractional habitability over all latitudes:

\begin{equation}
    f_{hab} = \int_P \int_{-\pi/2}^{\pi/2} \frac{H[d(t),\lambda] \, cos(\lambda)}{2P} \, d\lambda \, dt 
\end{equation}

We are primarily interested in quantifying the fraction of temporal habitability for planets orbiting stars of different stellar types, and used aqua planets as a test bed for observing general climate trends for varying eccentricity. Broadband planetary albedos (used as inputs to the EBM) increase monotonically with rising stellar effective temperatures \citep{Kasting1993, Selsis2007, Shields2013}. In our EBM, as the surface temperatures fall below $-2^{\circ} \, C$, ice forms and the broadband albedo changes correspondingly. We run the EBM with ``warm start" (starting from a climate similar to modern-day Earth) and ``cold start" (starting from globally ice-covered) conditions and calculate the fractional habitability once the climate reaches equilibrium. The difference in fractional habitability between the two initial conditions is a measure of climate hysteresis\textemdash the dependence of the climate state on its history.

\subsection{Model Inputs}

High resolution broadband albedos for planets orbiting F2V star HD128167, K2V star HD22049 \citep{Segura2003}, G2V star The Sun \citep{Chance2010}, and M3V star AD Leo\footnote{The Virtual Planetary Laboratory Spectral Database. 〈http://vpl.astro.washington.edu/spectra/stellar/mstar.htm} \citep{Reid1995, Segura2005} were calculated in previous work, using the Spectral Mapping Atmospheric Radiative Transfer Model, or SMART \citep{Meadows1996}, assuming an Earth-like atmosphere and surfaces composed of ocean, land, and ice of different grain sizes \citep{Shields2013}. We employ the broadband albedos of planets with ocean-covered surfaces for our aqua planets, with frozen regions corresponding to a $50\%$ mixture of snow and blue marine ice weighted by the corresponding SED, which is normalized to 100\% of the modern solar constant (1360 W/m$^2$). For more details on this approach, see Shields \emph{et al.} (\citeyear{Shields2013}).

We run our models with an obliquity $\theta = 0^{\circ}, 45^{\circ}, 90^{\circ}$ and the Earth's longitude of periastron, or azimuthal obliquity, $\omega = 102.065 ^{\circ}$. This angle only affects the climate of planets with non-zero obliquity and is defined from the vernal equinox, thus already accounting for the precession angle. The XUV flux for the Sun is taken from Airapetian \emph{et al.} (\citeyear{Airapatian2017}), who constructed it with both the Solar Dynamic Observatory and the Flare Irradiance Spectral Model (FISM). The AD Leo XUV flux taken from Chadney \emph{et al.} (\citeyear{Chadney2015}) was constructed using a coronal model. Additionally, in our mass loss calculations we do not include stellar evolution (i.e., we keep the stellar luminosity constant). All stellar parameter inputs are summarized in Table 2.

The XUV fluxes for both the M- and G-dwarf stars at an average Earth-Sun distance of 1 AU are scaled to the varying orbital distance of the eccentric planet over the course of its year, and used as input to our EBM. We assume here that the XUV flux scales linearly with the bolometric luminosity. An absorption efficiency factor of $\epsilon = 0.1$ is applied in our calculations. For this choice of XUV fluxes, our planets are well within the energy-limited regime for all eccentricities below 0.97 and 0.95 for the M- and G-dwarf planets, respectively. Above these eccentricities the XUV flux at the periastron passage is large enough ($>$ $10^{4} \frac{erg}{cm^2 s}$)so that radiative recombination significantly inhibits the rate of mass loss \citep{Murray2008}.

\subsection{Model Validation}

The EBM with broadband albedos as input from SMART was previously validated and shown to reproduce the Earth's current ice line latitude and global mean surface temperature to within 6$^\circ$ and 3$^\circ$\textbf{C}, respectively \citep{Shields2013}. Here we have validated the EBM with our RGH parameterization by reproducing the moist and RGH instellation thresholds of Wolf and Toon (\citeyear{Wolf2015}). As shown in Figure 2, the climate of our simulated G-dwarf planet with zero eccentricity remains stable up to $119\% \, S_0$\textemdash where $S_0$ is the modern solar constant for the Earth\textemdash compared to $121\% \, S_0$ found by \cite{Wolf2015}. The onset of the moist greenhouse and significant water loss occurs here at $116\% \, S_0$, compared to $119\% \, S_0$ in their study. Figure 3 shows the global mean temperature and the mean OLR as calculated with our EBM with the RGH parameterization, compared with the 3D CAM4 GCM (global climate model), and with the EBM with the traditional linear OLR parameterization. Our OLR parameterization produces a much better agreement in surface temperature with the GCM than the EBM with the default OLR parameterization, while not greatly changing the average OLR behavior as instellation increases. The CAM4 simulations exhibit a sharp increase in surface temperature and the mean OLR as the planet transitions into the moist greenhouse, once the solar constant is increased by $12.5\%$. However, as the solar constant increases, the climate is stabilized by the increasing top-of-atmosphere albedo, due to the formation of thick cloud decks \citep{Wolf2015}. The EBM does not include moist physics, and our RGH parameterization leads to a thermal runaway that is exponential with increasing instellation.

\section{Results}
\subsection{Fractional Habitability}

Figure 4 shows the warm start results in the eccentricity-instellation parameter space for all 4 stellar types. In our EBM, the cold edge of habitability is ultimately determined by the large ice-cap instability, which causes rapid collapse of the ice caps to the equator once the instellation falls below a certain threshold. On the warm end, habitability is truncated by the thermal runaway of the atmosphere. We find that in the case of a G-dwarf planet with an eccentricity $e = 0$, and an Earth-like atmosphere, the inner edge of EHZ corresponds to a stellar flux of $119\% \, S_0$, while the outer edge corresponds to $82.5\% \, S_0$, although with a significant ice cap. In the case of the M-, K-, and F-dwarf spectral types the outer and inner edges of the EHZ are $[70\% \, S_0 , \, 107.5\% \, S_0]$, $[80\% \, S_0 , \, 117\% \, S_0]$, $[84.5\% \, S_0 , \, 121.5\% S_0]$, respectively. With increasing eccentricity, planets with habitable conditions shift towards lower instellation values. On the K-, G-, and F-dwarf planets, the warming effects of eccentricity have a stronger impact on the inner rather than outer edge of the EHZ, due to the extra energy required to thaw sea ice on an ice-covered planet, compared with the transition from a water world to a moist hot house. On planets orbiting just outside of their host stars' full-orbit EHZ\textemdash the eccentricity-instellation parameter space over which any portion of the planet's surface is habitable throughout its entire year\textemdash a minimum eccentricity of about 0.2 is required to actually thaw sea ice. On the inner edge, we see a much steeper outward migration with increasing eccentricity.

In our set of modeled planets, we observe temporal habitability around any star with sufficient orbital eccentricity. A a sharp transition is seen in Figure 4, from planets with fractional habitability close to unity and exhibiting full-orbit habitability (yellow region), to planets with fractional habitability below 0.5 (light blue region). This light blue region consists of planets which experience globally frozen conditions for a fraction of the year and habitable conditions during the rest of the orbit. Cooler stars exhibit temporal habitability over a larger eccentricity-instellation parameter space. The region of temporal habitability on M-dwarf planets is $27\%, \, 34 \%$ and $39 \%$ larger than their K-, G-, and F-dwarf analogs, respectively. Moreover, the minimum eccentricity and instellation required for the appearance of temporal habitability is smaller for planets orbiting cooler stars. For our M-dwarf planets, we observe temporal habitability at an eccentricity as low as $e = 0.13$. The minimum eccentricity required for the appearance of temporal habitability around the K-, G-, and F-dwarf planets is $0.230$, $0.270$, and $0.285$, respectively . For larger eccentricities, temporal habitability appears over a larger range of instellations for any host star. At the same time, the instellation range of the full-orbit EHZ shrinks with increasing eccentricities. At eccentricities above $e = 0.6$, the region of temporal habitability becomes a significant component of the total EHZ and constitutes $\sim$50\% of the total at $e = 0.8$.

The difference in the response of the inner and outer edges is responsible for the gradual decline and shift of the EHZ to lower instellations up to $e = 0.5$, where the appearance of temporal habitability expands the outer habitable edge (Figure 5). Due to the appearance of temporal habitability at eccentricities as low as $\sim$ 0.13, the M-dwarf planet has no decline in the EHZ until all planets enter the RGH at extreme eccentricities.

\subsection{Mass Loss and Runaway Greenhouse Planets}
We calculate the annual mass loss rates for the Sun and for the mid-type M-dwarf star, AD Leo. Figure 6 shows the time (in Myr) it takes to lose Earth's entire surface water inventory as a function of eccentricity and instellation. For planets in a RGH with the same eccentricity and instellation, the mass loss rate is directly proportional to the XUV flux (see equation 3). Due to the fact that AD Leo produces $\sim$6x larger XUV flux than the Sun, the water loss is $\sim$6x larger. We find that the eccentricity threshold for thermal runaway is lower on the M-dwarf planet by $12 \% \,  S_0$at $e = 0$ and $5.5 \% \, S_0$ at $e = 0.9$ relative to the G-dwarf planet.

On a circular orbit, a G- and an M-dwarf planet in a RGH receiving $120\% S_0$ can become desiccated in $\sim$ 3.6 Gyr and $\sim$ 690 Myr, respectively. We find that at any eccentricity the mass loss becomes significant at a few percent of $S_0$ below the thermal runaway threshold. For the G-dwarf planet at $e = 0$ and instellations between $118.5 \%$ and $119.5 \% \ S_0$, prior to the thermal runaway the planetary conditions are conducive to the loss of Earth's surface water inventory in $7.3 - 5.2$ Gyr. Similarly, the M-dwarf planet receiving $106.5\% - 107.5\%S_0$ (thermal runaway occurs with a flux of $108\% S_0$) has desiccation timescales of $1.32 - 1.03$ Gyr. This region of significant water loss while in the moist greenhouse exhibits itself as the small strip immediately to the left of the black contour indicating the transition from the moist to RGH region. Across the $0 - 0.9$ eccentricity range, the annual water loss rate in a RGH state varies by a factor of $\sim$2 at most, regardless of the host star SED.

\subsection{Bistability}
Figure 7 shows the fractional habitability on M-, K-, G-, and F-dwarf planets at varying eccentricity assuming cold start initialization. A comparison with the warm start results shown in Figure 4 reveals two different outcomes, depending on starting conditions\textemdash a situation we refer to here as ``bistability". On circular orbits, planets orbiting all stellar types exhibit bistability in some instellation range. The cooler the host star, the smaller this range of bistability. As shown in Figure 8, climate hysteresis (approximated by the warm minus cold start difference in fractional habitability) decreases with increasing eccentricity for any stellar type. In the case of the M-dwarf spectrum, planets with eccentricities above 0.26 exhibit no bistability. Similarly, in the K-,G-, and F-dwarf cases, no bistability (seen as the yellow wave-shaped region) occurs at eccentricities above 0.45, 0.48, and 0.5, respectively.

The outer edge of the full-orbit EHZ is affected by the initial conditions at any eccentricity. In both warm and cold start cases, planets only exhibit differences in habitability in the bistability region at low eccentricities or along the interface between the temporal and full-orbit habitable regions (seen as the long blue arc), indicating an expansion of the region of temporal habitability on cold start planets compared to those simulated with warm start conditions.

\subsection{Habitability at Higher Obliquities}
To test the sensitivity of our results to changes in obliquity, we repeated our zero-obliquity calculations of fractional habitability on warm start M-dwarf planets for larger obliquities ($45^{\circ}$ and $90^{\circ}$), as shown in Figure 9. Obliquity has a small effect on the inner edge of the EHZ, due to the absence of sea ice on these planets. Over the entire range of eccentricities, for an M-dwarf planet with an obliquity of $90^{\circ}$, the inner edge of the EHZ occurs at a flux that is at most 2.5\% larger than its $0^{\circ}$ obliquity counterpart. However, the outer edge of the habitable zone is more sensitive to changes in obliquity, due to the presence of ice caps (or ice belts) on these planets. The outer edge of the EHZ appears at a flux that is at most 7\% higher on planets with 90$^\circ$ compared to 0$^\circ$ obliquity. Planets with 90$^{\circ}$ obliquity also exhibit a region of small fractional habitability ($ < 0.1$), corresponding to habitable conditions at the poles, constituting both full-orbit (e $\approx$ 0) and temporarily habitable EHZ regions (e $>>$ 0). For planets with zero eccentricity, this region spans an instellation range of $68 - 76\% \, S_0$. With increasing eccentricity it extends outside of the full-orbit EHZ to include planets with eccentricities as large as 0.9. With this region of small fractional habitability included, planets with 90$^\circ$ obliquity have an EHZ that is 46$\%$ larger than their 0$^\circ$ obliquity counterparts.

We find that larger obliquity leads to warmer climates on planets in the full-orbit EHZ. With increasing obliquity the ice caps/belts retreat and fractional habitability approaches unity. For planets with a $90^{\circ}$ obliquity, we see no sea ice, and fractional habitability is equal to 1 within the full-orbit EHZ.

\section{Discussion}

In this work we used an EBM with a simple RGH parameterization to explore the effects of eccentricity, obliquity and host star SED on habitability and water loss of terrestrial aqua planets with Earth-like atmospheres orbiting F-, G-, K-, and M-dwarf stars. The instellation range over which planets exhibit habitable surface conditions throughout their entire orbit shrinks with increasing eccentricities, but the emergence of a temporarily habitable zone helps to compensate for this reduction of the full-orbit EHZ. The temporarily habitable zone widens with decreasing host star effective temperature. Uniquely for M-dwarf planets, the total EHZ (temporal + full-orbit) widens with increasing eccentricity up to e $\approx$ 0.6. For planets in a RGH, Earth's entire surface water inventory can be lost in a few Gyr. Similar water loss is achieved in a moist greenhouse state once water mixing ratios approach unity in the tropical regions of the planet. Earth-like aqua planets on eccentric orbits remain habitable during some portion of their orbits for a wide range of instellations, and reduced or eliminated bistability increases the likelihood that an observed eccentric planet is in a state determined purely by its current orbital configuration.

Given the sensitivity of the inner habitable-zone edge to increases in eccentricity, if an Earth-like aqua planet on a circular orbit were perturbed to a higher eccentricity it could potentially enter the RGH state, desiccating the surface. In the case of the G-dwarf planet receiving $100 \% \, S_0$ this occurs in Figure 6 at eccentricities above $e = 0.55$. For the M-dwarf planet receiving $100 \%  \, S_0$, the minimum eccentricity required for the thermal runaway of the planet is $e = 0.38$.

For any eccentricity, the EHZ shifts towards lower instellations for cooler stars, due to the lower albedos of surface ice and snow on orbiting planets, and the absorptive properties of atmospheric gases. At longer wavelengths ice and snow absorb more strongly \citep{Dunkle1956}, leading to smaller broadband planetary albedos \citep{Shields2013, Shields2014}, more efficient thawing of sea ice, and a wider region of temporal habitability on planets orbiting cooler stars. Additionally, greenhouse gases like water vapor and CO$_2$ absorb more strongly in the IR, which leads to a thermal runaway of the climate at lower values of instellation. Sea ice thaws more efficiently on planets orbiting cooler stars. While this holds true for planets at any eccentricity, the differences in the fractional habitability due to different spectral energy distributions diminish with increasing eccentricities\textemdash at high eccentricities different stars have similar fractional habitability as a function of instellation.

We find that increasing obliquity shrinks the EHZ for planets at any eccentricity. Increasing obliquity leads to an inward migration of the outer edge of the EHZ, due to ice-albedo feedback and ice-sheet instability. These mechanisms do not affect planets near the inner edge of the habitable zone due to the total absence of sea ice. This effect is strongest at e = 0. With increasing eccentricity, the increasing seasonal variations in radiative forcing dominate the effects of obliquity as with the effects of stellar spectral energy distribution. We find that larger obliquity warms the climates of planets within the full-orbit EHZ. However, at higher obliquities the sea ice becomes increasingly unstable, leading to larger climate hysteresis.

For our choice of orbital period (360 days), we find no planets exhibiting periodic phases of RGH conditions at any eccentricity. This fact should hold true for any planet with smaller orbital periods. However, with larger periods, for a sufficiently large eccentric orbit, planets may be able to spend enough time at or near apoastron to cool sufficiently for evaporated oceans to condense back onto the surface. Additionally, the moist greenhouse region in this parameter space is quite small in both the M- and G-dwarf cases, but the large water loss rates achieved for moist greenhouse planets may challenge the habitability of these planets.

In our mass loss calculations we assume a constant, quiescent XUV flux. While this may be sufficient for older and more dormant main-sequence stars, the flare activity on a mid-type M-dwarf such as AD Leo may alter the physics of atmospheric escape. Additionally, for XUV fluxes exceeding $\sim$ 0.4 W/m$^2$ the absorption efficiency decreases rapidly\citep{Bolmont2017}. In the case of AD Leo, this flux is achieved at periastron passage at an eccentricity of 0.73. In the case of the Sun, such flux values are only achieved at eccentricities above 0.89. Our water loss timescales may therefore be overestimated on planets with the largest eccentricities for an assumed constant absorption efficiency factor of 0.1.

We find a large drop in fractional habitability along the transition from full-orbit to temporarily habitable planets. This sharp transition is likely due to a combination of effects. The ice-albedo feedback and the ice-cap instability accelerate the expansion of the ice caps as planets move towards apoastron. Planets that freeze over remain frozen for a significant amount of time, significantly reducing their fractional habitability. This causes a sharp transition between completely thawed planets and those that are temporarily frozen.

Our choice of constant water mixing ratios in the three temperature brackets may lead to over or underestimation of water loss of order unity on moist greenhouse planets, depending on the instellation. However, owing to the fact that water mixing ratios rise rapidly above 340K, this approximation is appropriate for determining the mass loss limited inner boundary of the EHZ to within one percent of the solar constant.

The EHZ assumes an Earth-like atmosphere, with fixed CO$_2$. Planets on eccentric orbits may experience changes in atmospheric CO$_2$ concentration as surface temperatures vary throughout the orbit if a CO$_2$ cycle operates on these planets, as it does on the Earth \citep{Walker1981} and Mars \citep{Hess1980}. Including a dynamic, orbital distance-driven CO$_2$ cycle into a climate model would be useful towards further refining the boundaries of the EHZ.

We assume constant broadband planetary albedos throughout our simulations, given an Earth-like atmosphere. The reflective properties of a planet's atmosphere will likely change with temperature, water vapor mixing ratio, and CO$_2$ concentration, affecting the planet's overall (atmosphere + surface) broadband albedo. Incorporating into a climate model a temperature-dependent broadband planetary albedo parameterization that accounts for variations in the atmospheric concentration throughout the temperate, moist, and RGH regimes would be an important step to take in future work.

\section{Conclusions}
Using a 1-D EBM with a simple runaway greenhouse parameterization, we have demonstrated that eccentric planets orbiting cooler stars exhibit temporal habitability in a larger region of the eccentricity-instellation parameter space compared with planets orbiting hotter, more luminous stars. This difference is largely due to lower relative ice and snow surface albedos, leading to more efficient thawing of sea ice on planets orbiting cooler, redder stars. Our approach reveals a refined eccentric habitable zone (EHZ) that is sensitive to host star SED and planetary obliquity. Additionally, our runaway greenhouse parameterization allowed us to calculate the inner boundary of the eccentric habitable zone (EHZ) with much greater accuracy than the traditional linear OLR parameterization. Orbital eccentricity leads to a rapid outward migration of the inner edge of the EHZ and a slower outward migration of the outer edge, leading to an overall reduction of the EHZ for all our simulated planets except M-dwarf planets. However, this reduction is somewhat alleviated at eccentricities above $\sim$ 0.5 by the appearance of a sizable temporal habitable zone. Conversely, the EHZ on eccentric planets orbiting M-dwarf stars widens with increasing eccentricity until all planets enter a runaway greenhouse state at extreme eccentricities. While in a runaway greenhouse, the M-dwarf planet experiences 6$\times$ greater water loss compared to its G-dwarf counterpart. Across the $e = 0 - 0.9$ eccentricity range the water loss rates in a runaway greenhouse state vary by a factor of 2. We also find that increasing planetary obliquity shrinks the EHZ, due to the inward migration of the outer edge of the EHZ, at the same time warming the climate of the planets in the full-orbit EHZ. Our study of bistability, through a fractional habitability comparison, reveals that the climates of planets with non-zero orbital eccentricities may be less sensitive to their histories. Bistability disappears all together with eccentricities larger than 0.26 on M-dwarf planets, an eccentricity much smaller than the 0.46 - 0.5 values required for no bistability on planets orbiting hotter K-, G-, and F-dwarf stars.

\section{Acknowledgments} \label{sec:acknowl}
This material is based upon work supported by the National Science Foundation under Award No. 1753373, and by a Clare Boothe Luce Professorship. We would like to acknowledge the high-performance computing support from Cheyenne (doi:10.5065/D6RX99HX) provided by NCAR's Computational and Information Systems Laboratory, sponsored by the National Science Foundation, and the high-performance computing cluster at the Research Cyber infrastructure Center  (RCIC), which provides systems, application software, and scalable storage support to the UCI research community. A.S. thanks the Virtual Planetary Laboratory at the University of Washington for nurturing the types of long-term collaborations that resulted in this paper. We thank Eric Wolf for providing data for comparison in Figure 3, and Cecilia Bitz for the original coding of the North and Coakley model.


\begin{table}[h!]
\centering
\begin{tabular}{ |c|c|c| }
 \hline
 n & T(K) & mixing ratio    \\
 \hline
 0 & $<$ 340     & 0               \\
 1 & 340 - 350  & $3\cdot 10^{-3}$\\
 2 & 350 - 370   & $10^{-1}$         \\
 3 & $>$ 370     & 1               \\
 \hline
\end{tabular}
\caption{The temperature regimes and corresponding average water mixing ratios used in our simulations with an EBM with a runaway greenhouse parameterization.}
\end{table}

\begin{table}[h!]
\centering
\begin{tabular}{ |c|c|c|c|c| } 
 \hline
 Star & \multicolumn{3}{|c|}{Broadband Albedos} & $F_{XUV}(a=1AU)$ \\
 \hline
 & Land Albedo & Ocean Albedo & Ice Albedo & $mW/m^2$\\
 \hline
 AD Leo (M3V)    & 0.332 & 0.234 & 0.315 & 29.4\\
 HD 22049 (K2V)  & 0.401 & 0.302 & 0.401 & --- \\
 The Sun (G2V)   & 0.415 & 0.319 & 0.514 & 5.60\\
 HD 128167 (F2V) & 0.414 & 0.329 & 0.537 & --- \\
 \hline
\end{tabular}
\caption{Stellar and planetary parameters used as input to the EBM, including broadband planetary albedos from Shields \emph{et. al} (\citeyear{Shields2013}) and incident XUV fluxes from Airapaitian \emph{et. al} (\citeyear{Airapatian2017}) and Chadney \emph{et al.} (\citeyear{Chadney2015}).}
\end{table}

\begin{figure}
  \centering
      \includegraphics[width=0.8\textwidth]{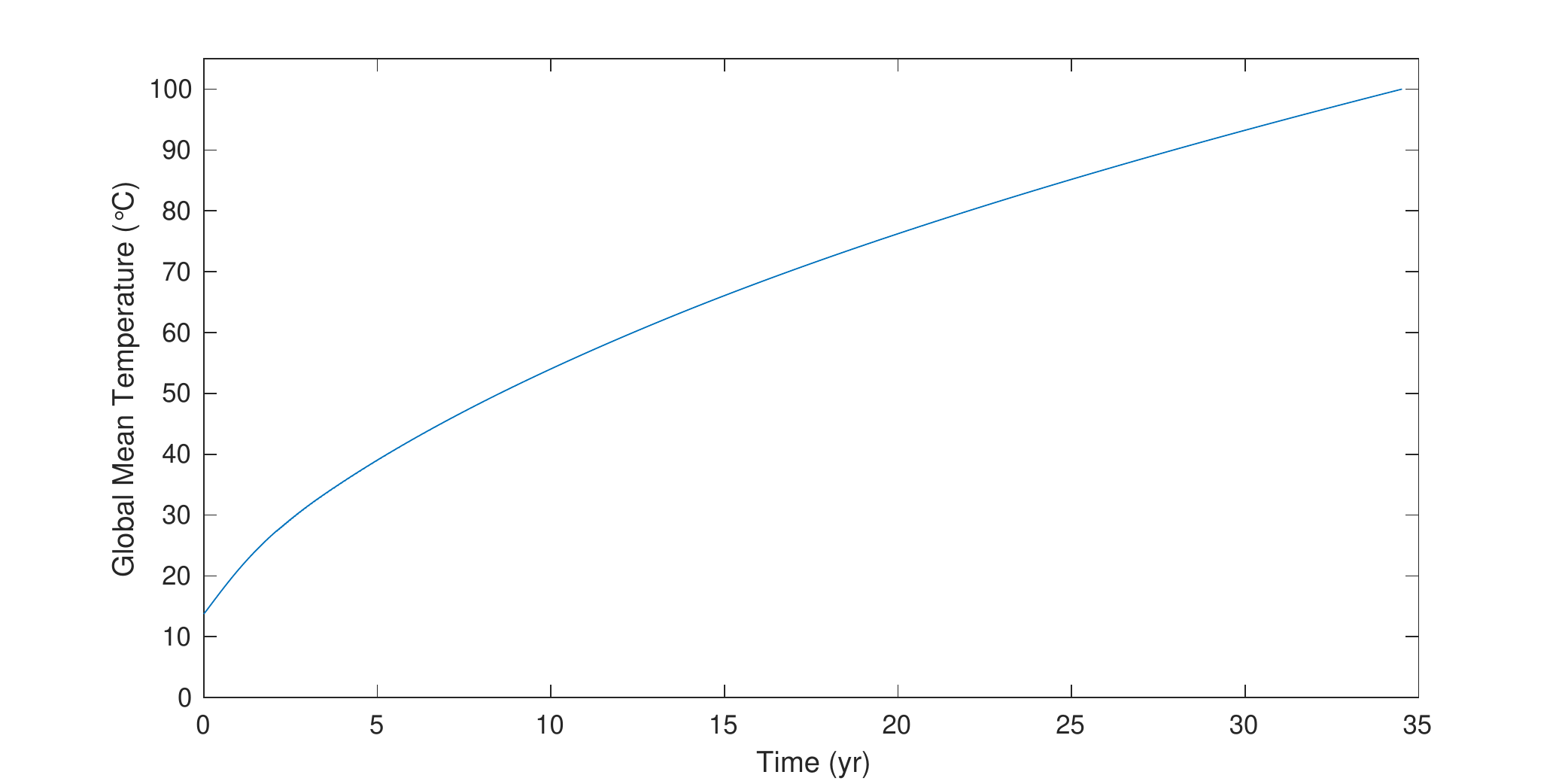}\\
      \includegraphics[width=0.8\textwidth]{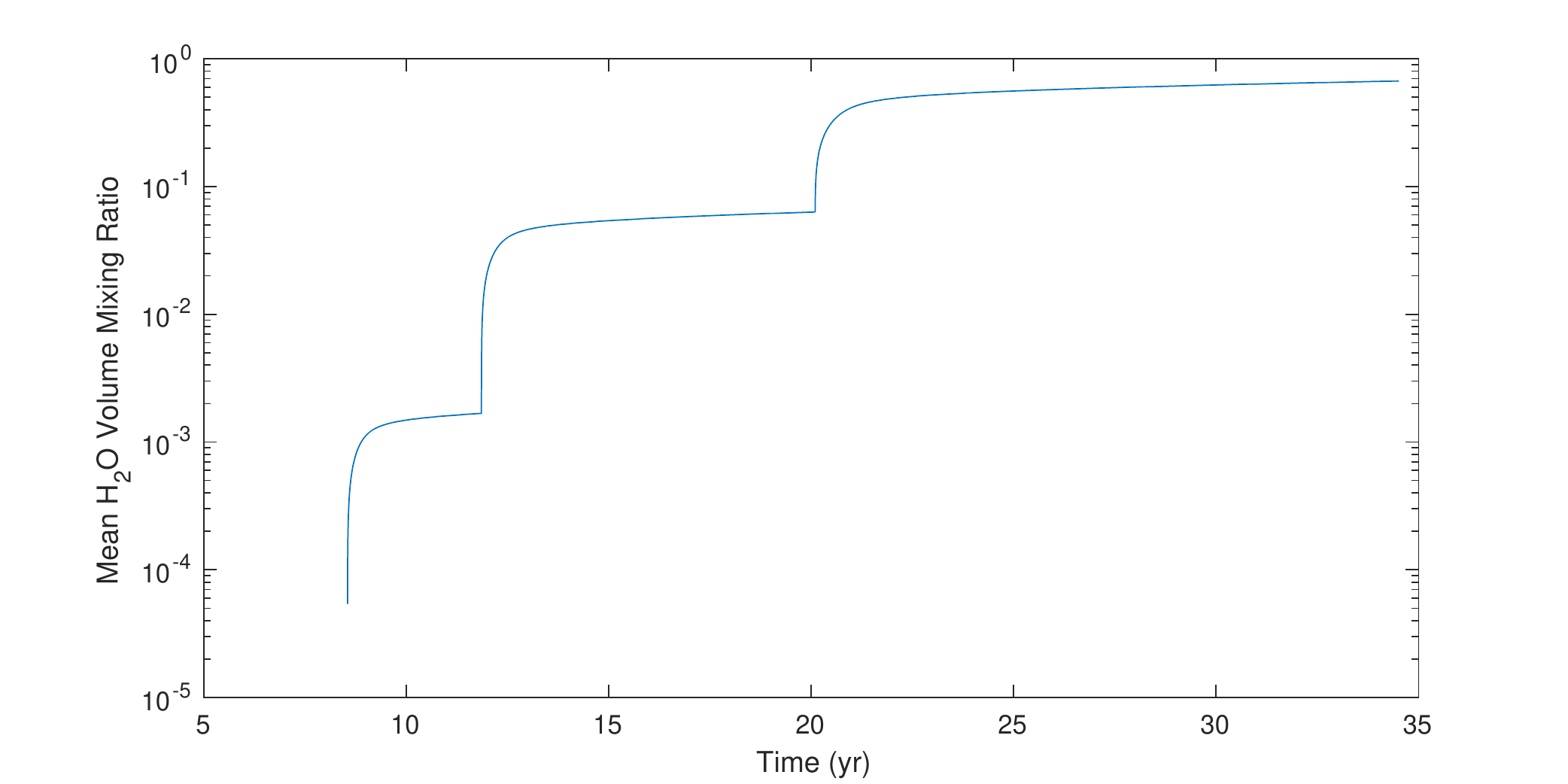}\\
      \includegraphics[width=0.8\textwidth]{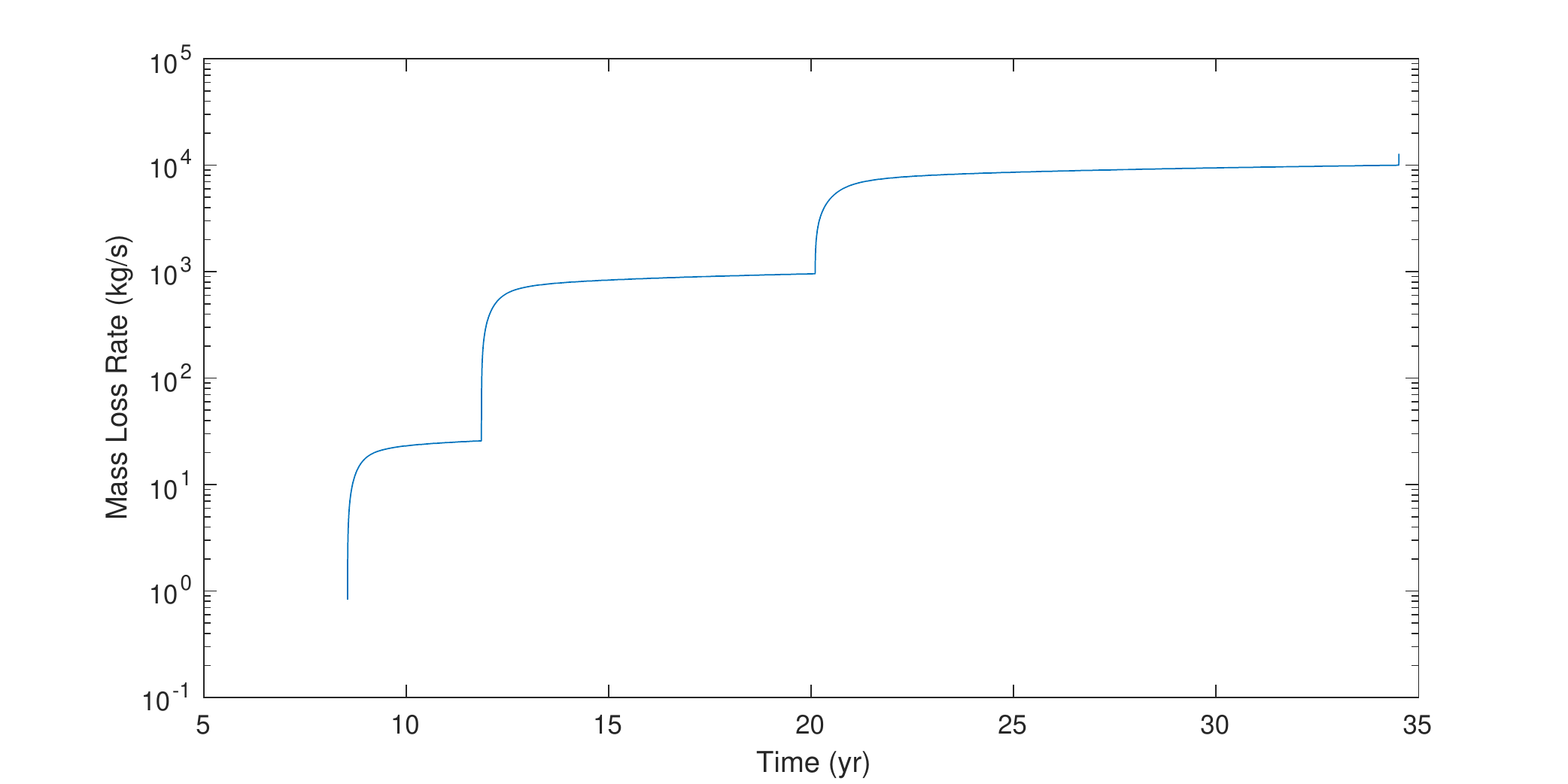}
  \caption{The evolution to the equilibrium state of the G-dwarf planet with e = 0 receiving receiving $125\% S_0$, where $S_0$ is the modern solar constant (the instellation on modern day Earth): Mean global temperature (top), mean water mixing ratio (middle) and mass loss rate (bottom) .}
\end{figure}

\begin{figure}
  \centering
      \includegraphics[width=0.8\textwidth]{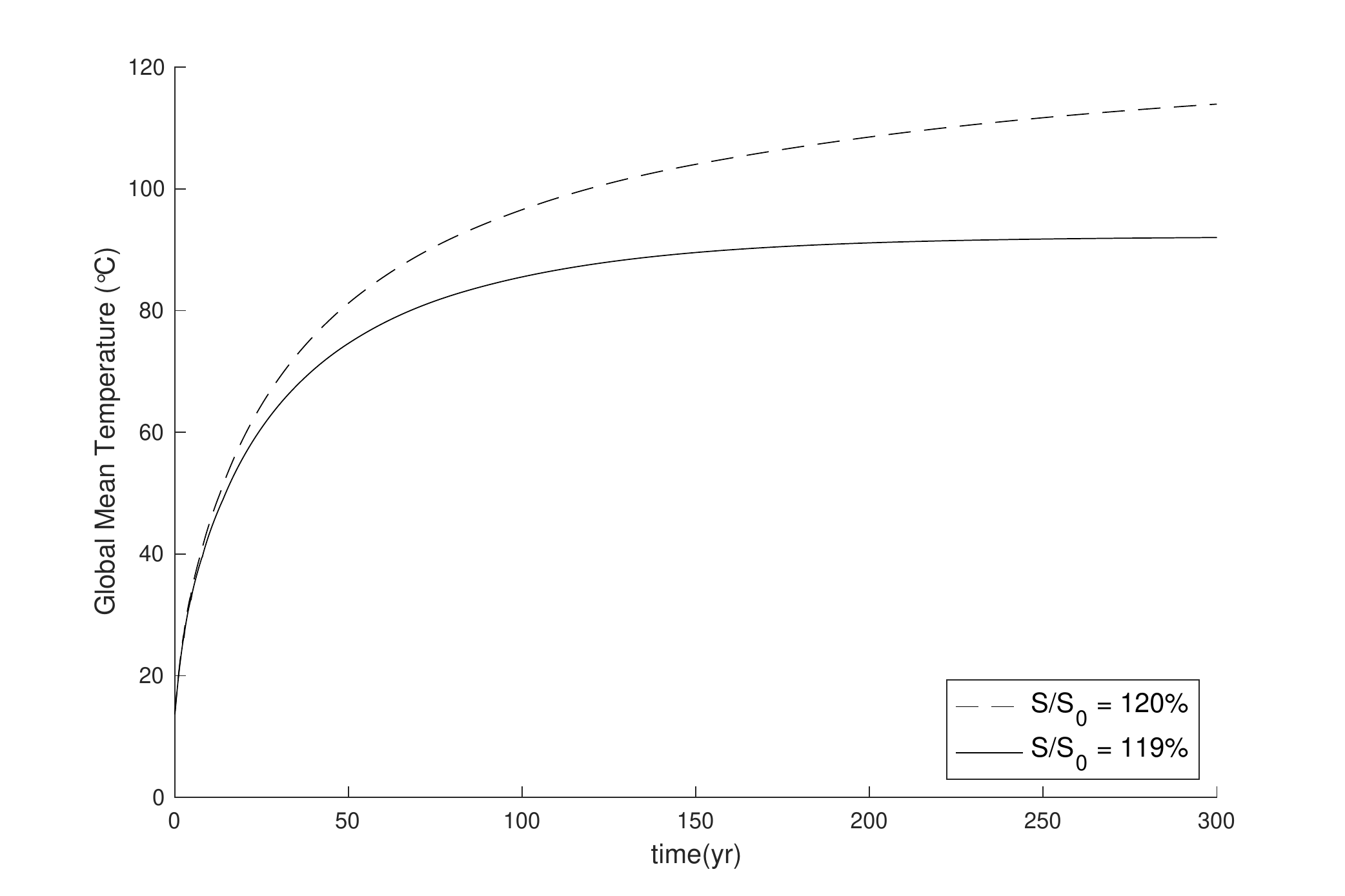}
  \caption{The evolution of global mean surface temperature on a G-dwarf planet with e = 0, receiving $119\% S_0$ (solid line) and $120\% S_0$ (dashed line), where $S_0$ is the modern solar constant. The latter planet receives the minimum instellation required for a thermal runaway.}
\end{figure}

\begin{figure}
  \centering
      \includegraphics[width=0.8\textwidth]{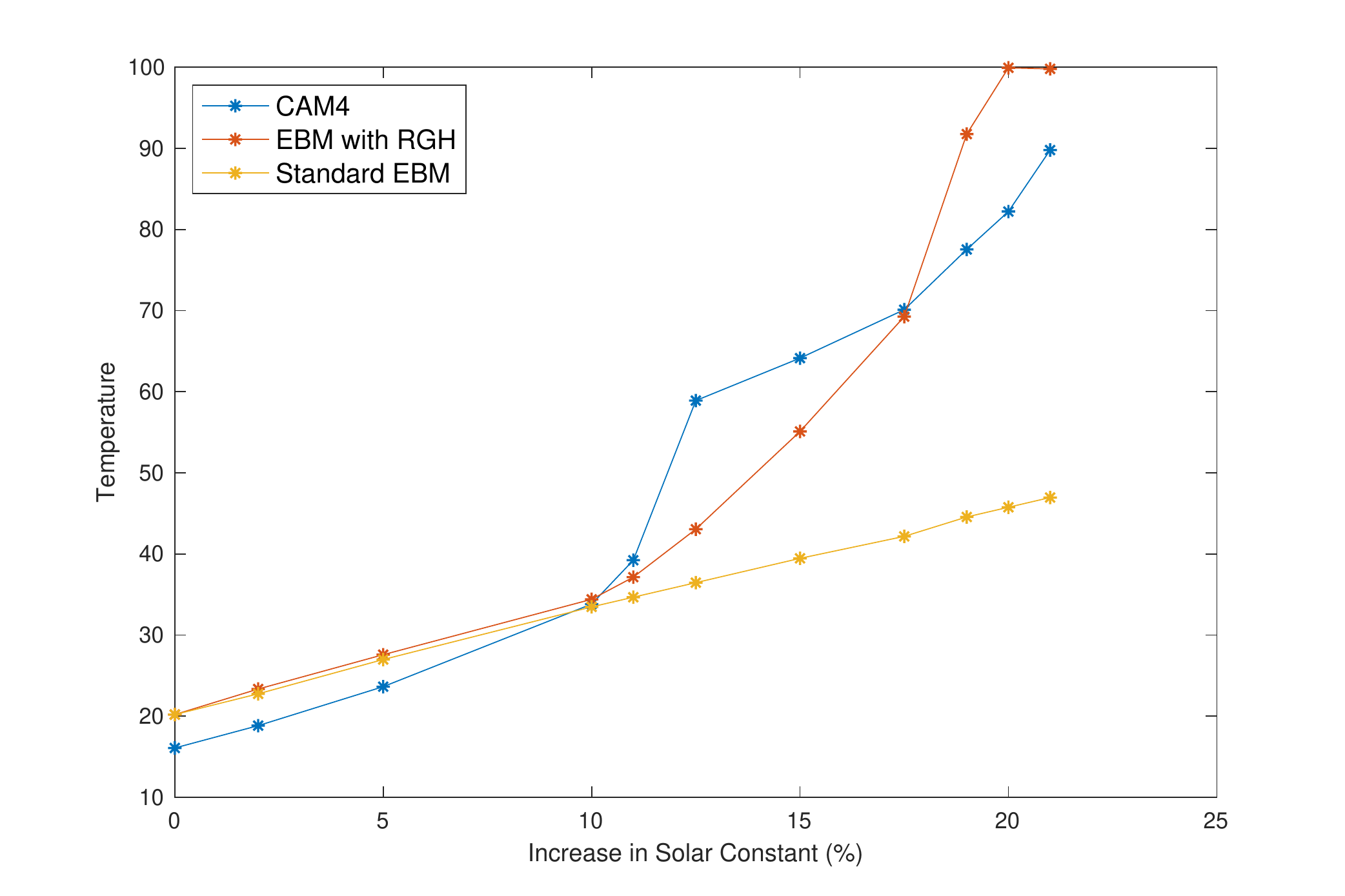}\\
      \includegraphics[width=0.8\textwidth]{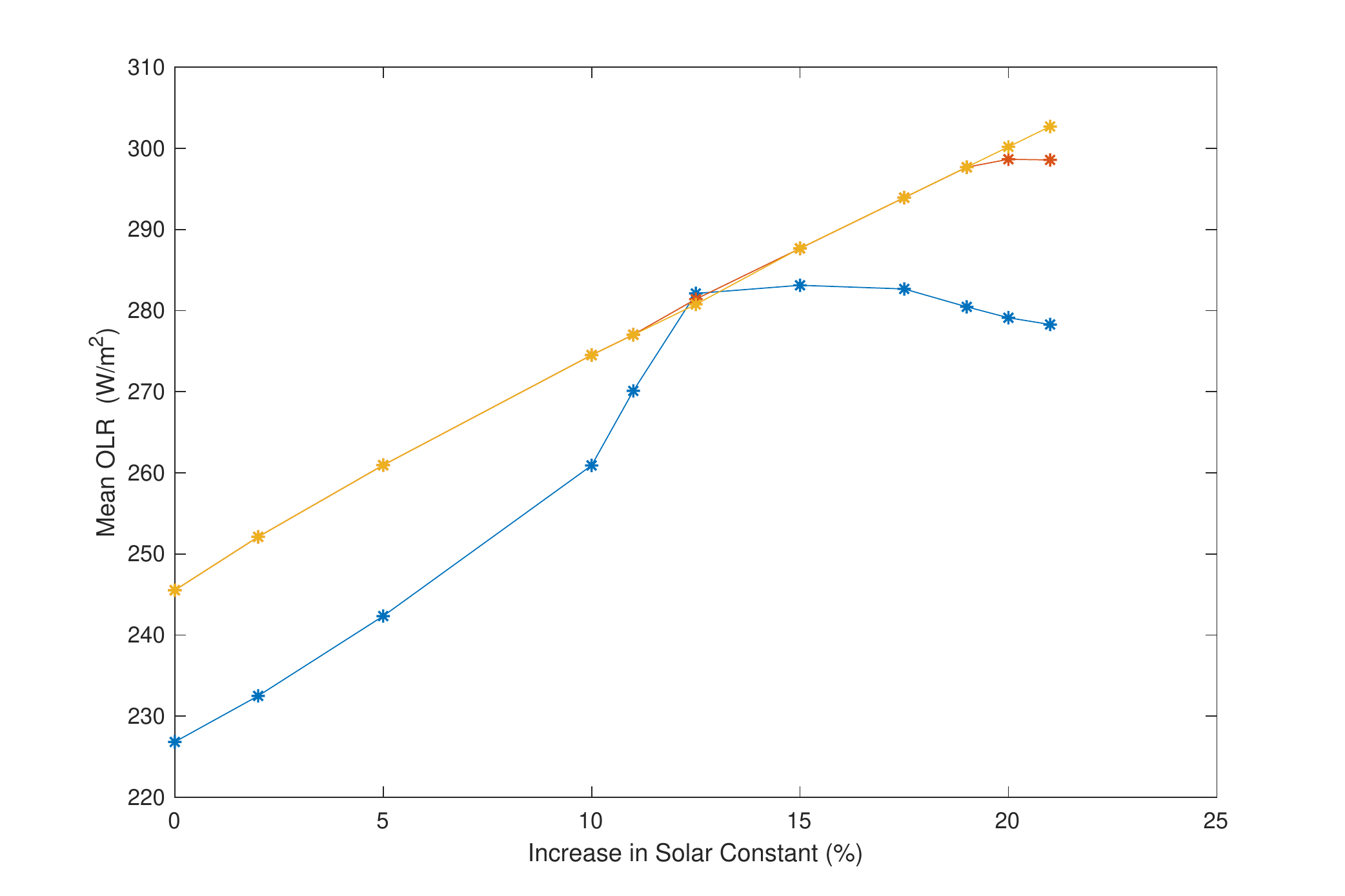}
  \caption{Top: Global mean surface temperature vs. increase in instellation in percent of the modern solar constant for a G-dwarf planet, using the standard EBM, the EBM modified to include a RGH parameterization, and the CAM4 GCM. Bottom: Comparison of the mean OLR between the three models. The CAM4 data is from Wolf and Toon \emph{et al.} (\citeyear{Wolf2015})}. 
\end{figure}

\begin{figure}
\begin{center}
\includegraphics [scale=0.42]{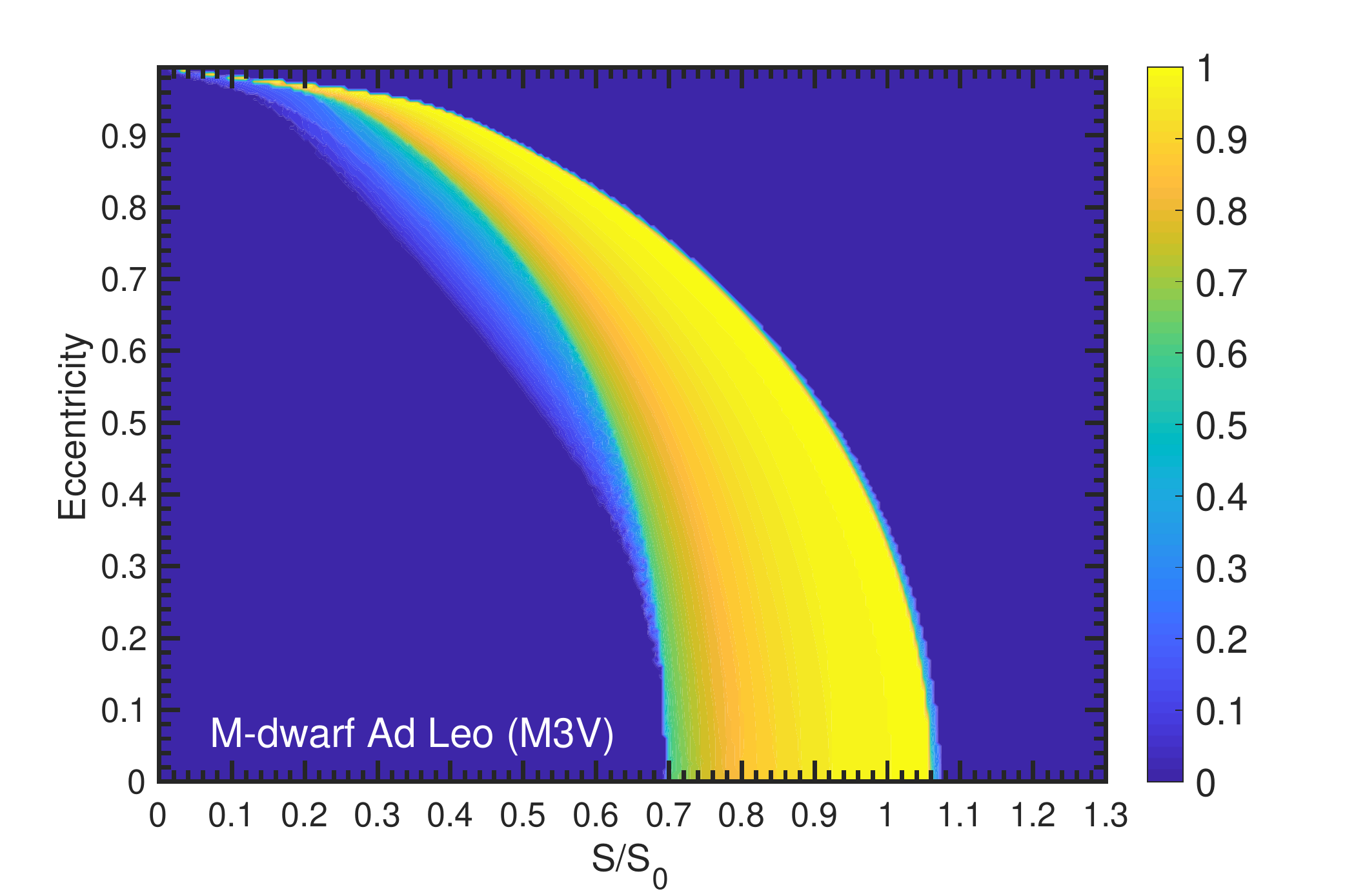}
\includegraphics [scale=0.42]{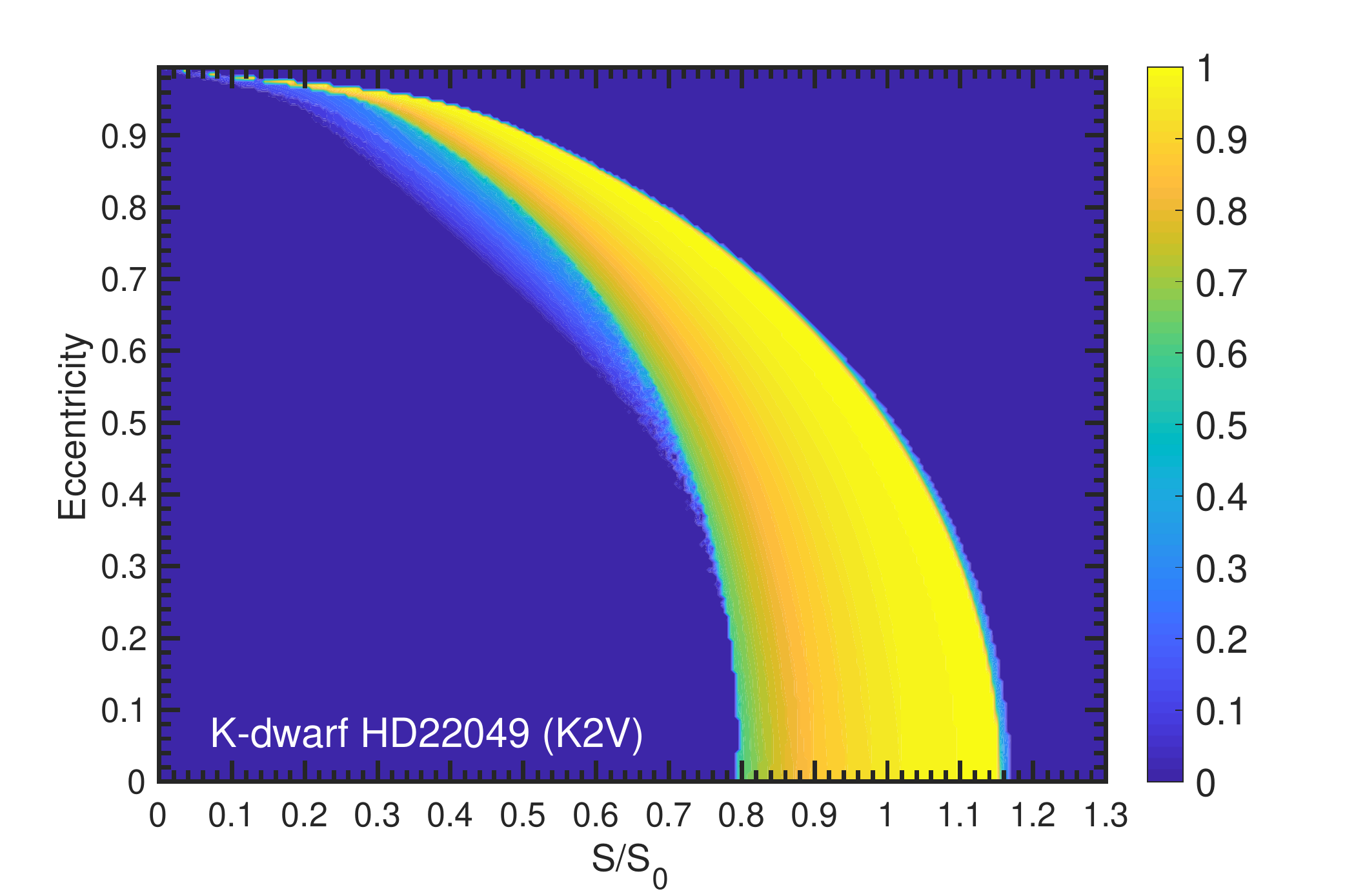}\\
\includegraphics [scale=0.42]{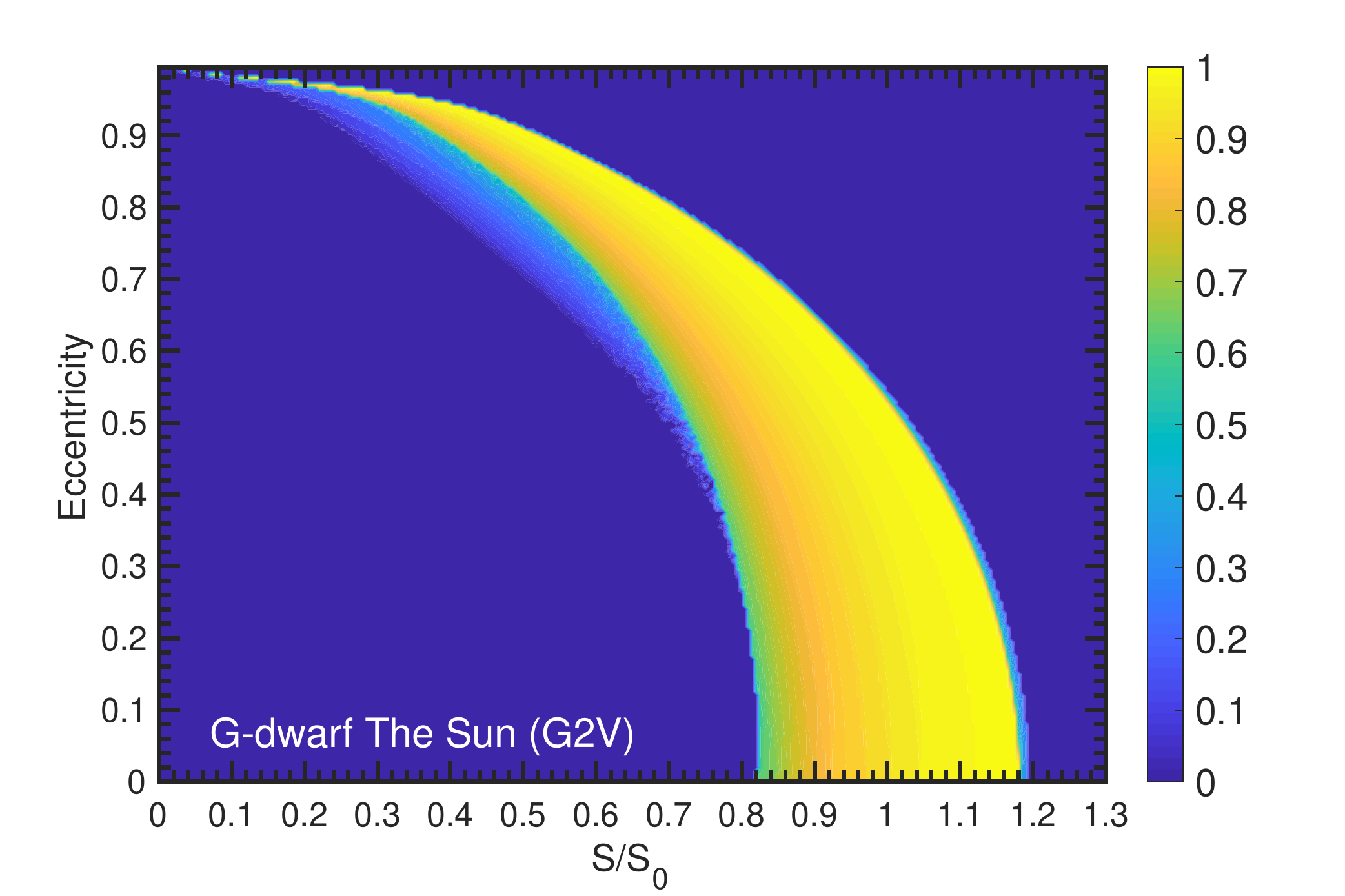}
\includegraphics [scale=0.42]{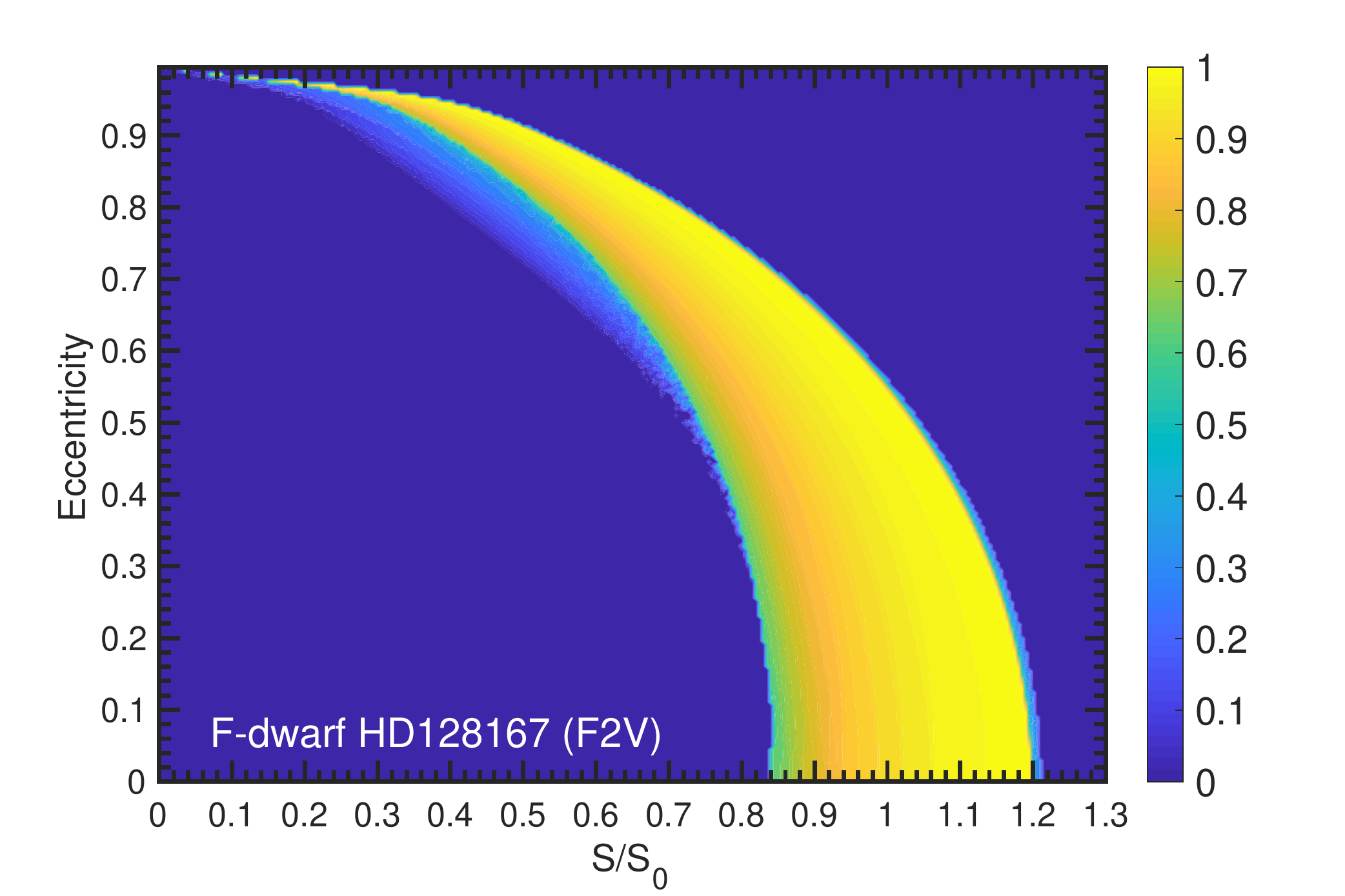}\\
\caption{Fractional Habitablity on the M-,K-,G- and F-dwarf planets with varying eccentricity and instellation, after EBM simulations assuming warm start conditions.}
\end{center}
\end{figure}

\begin{figure}
  \centering
      \includegraphics[width=0.8\textwidth]{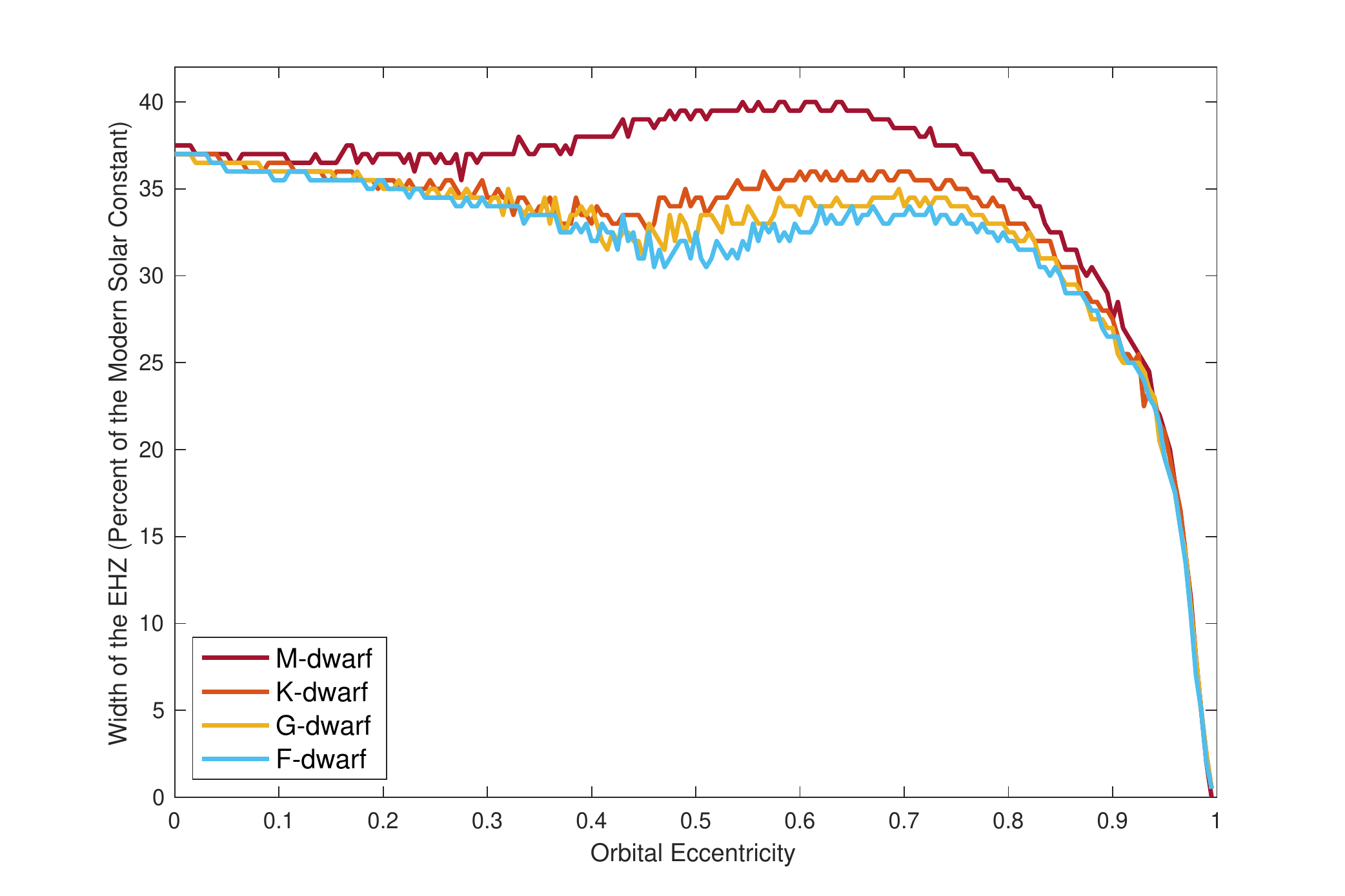}
  \caption{The range of instellations for which warm start planets orbiting M-,K-,G- and F-dwarf stars exhibit \textbf{(non-zero)} fractional habitability, as a function of orbital eccentricity.}
\end{figure}

\begin{figure}
  \centering
      \includegraphics[width=0.85\textwidth]{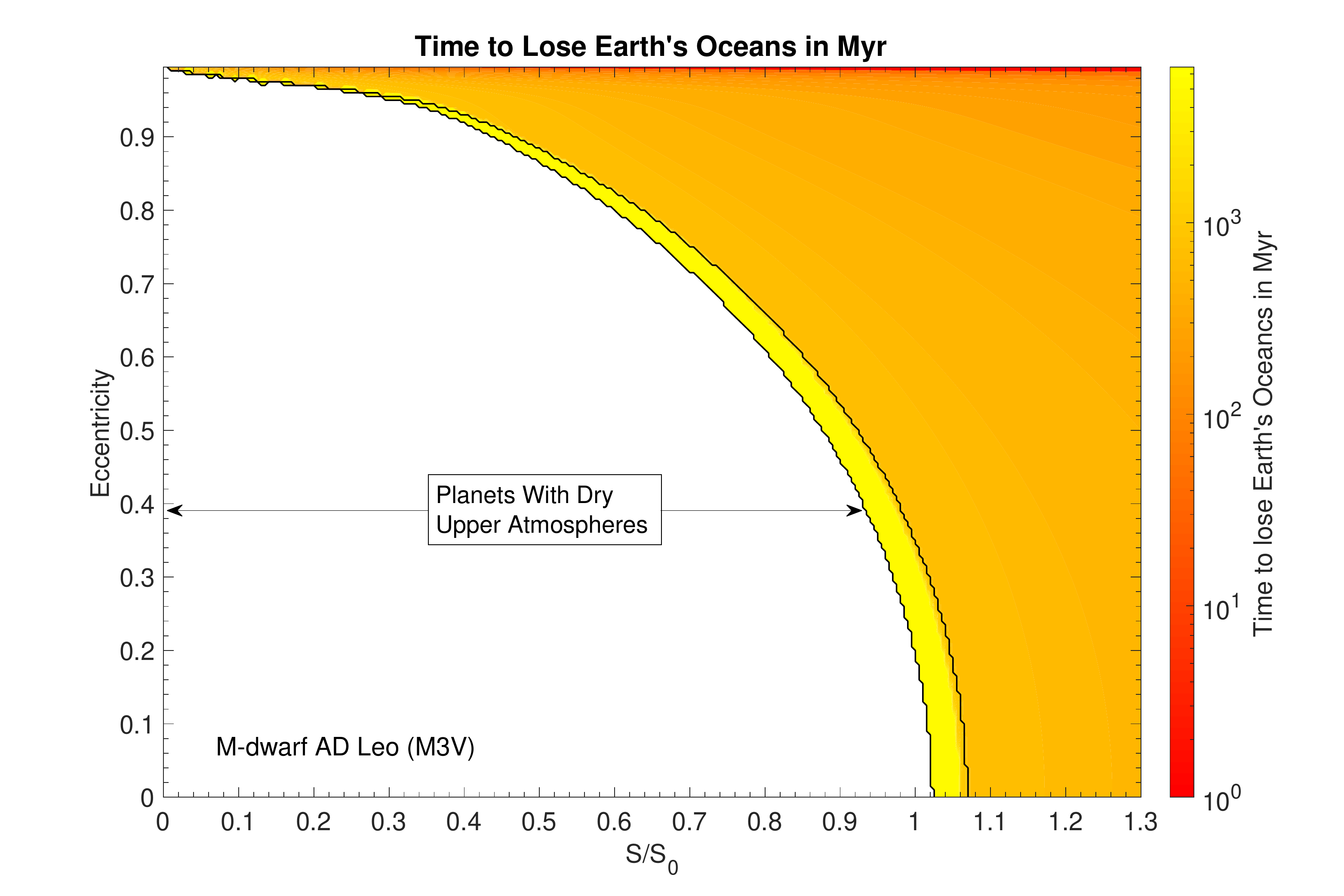}
      \includegraphics[width=0.85\textwidth]{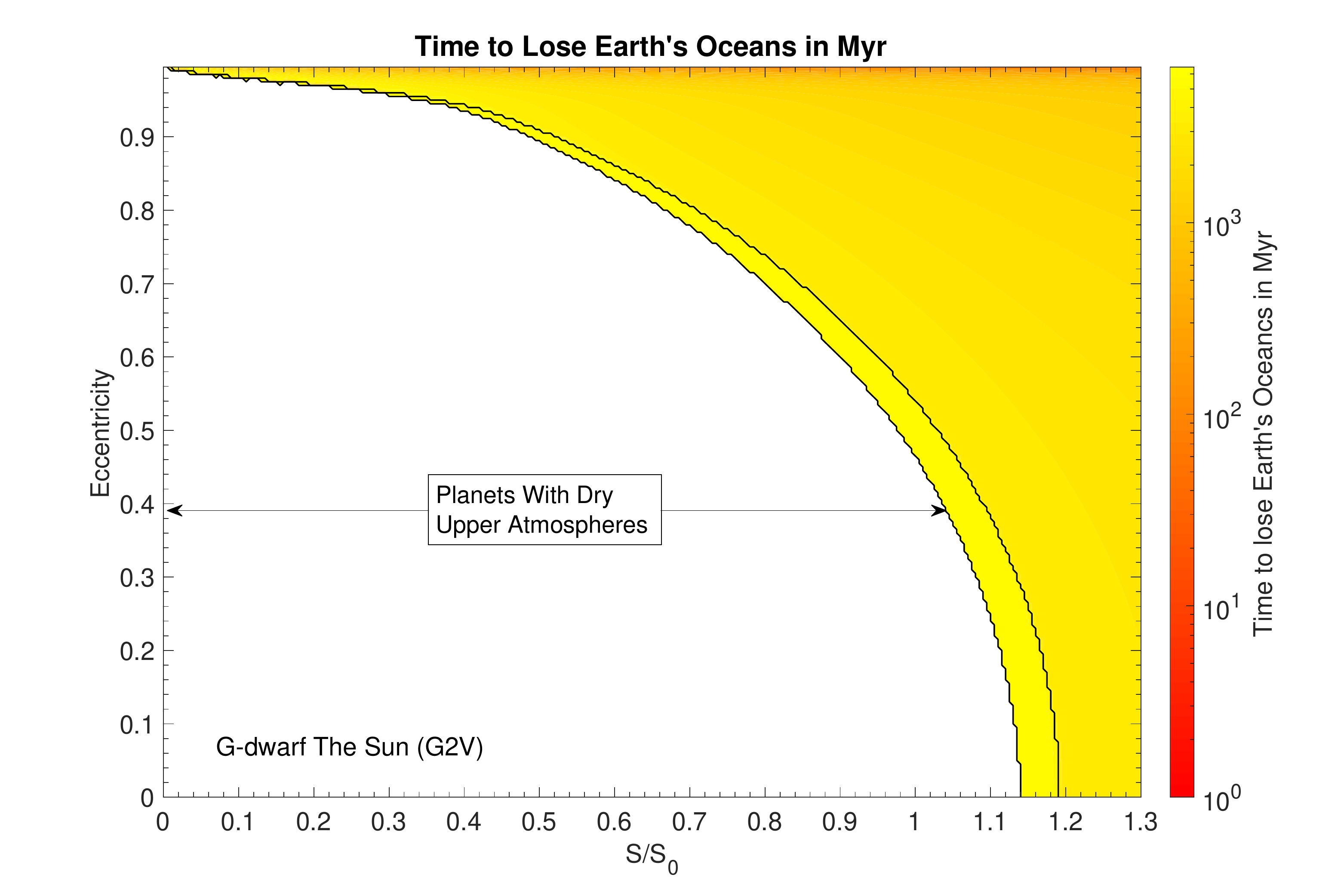}
  \caption{Time to lose Earth's entire surface water inventory on aqua-planets orbiting M-dwarf AD Leo and G-dwarf star The Sun in Myr. The M-dwarf planet is exposed to $\sim$6x more XUV flux than the G-dwarf planet with an equivalent climate, leading to a $\sim$6x higher mass loss rate. The black contour outlines the boundaries of the moist greenhouse.}
\end{figure}

\begin{figure}
\begin{center}
\includegraphics [scale=0.42]{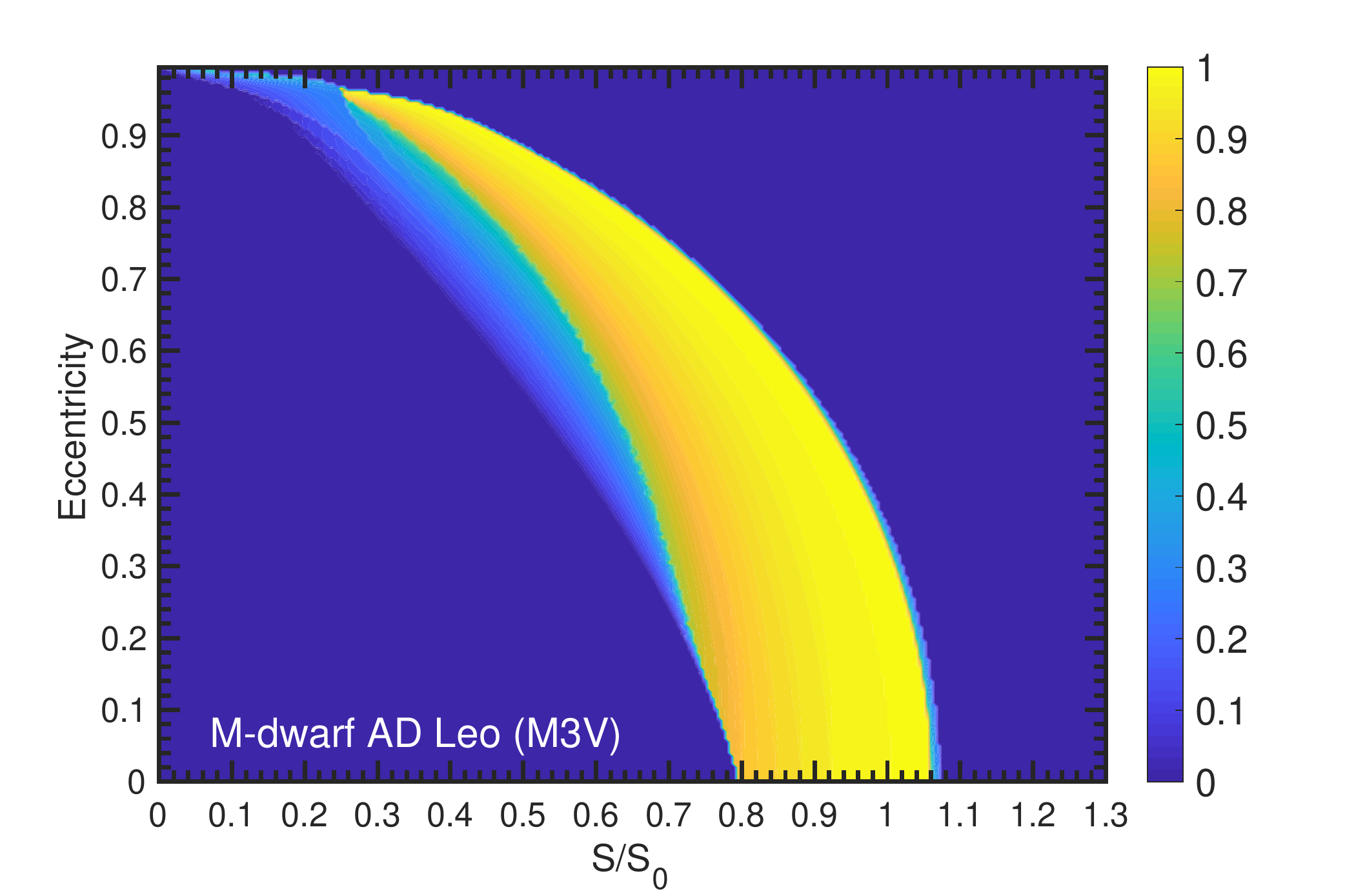}
\includegraphics [scale=0.42]{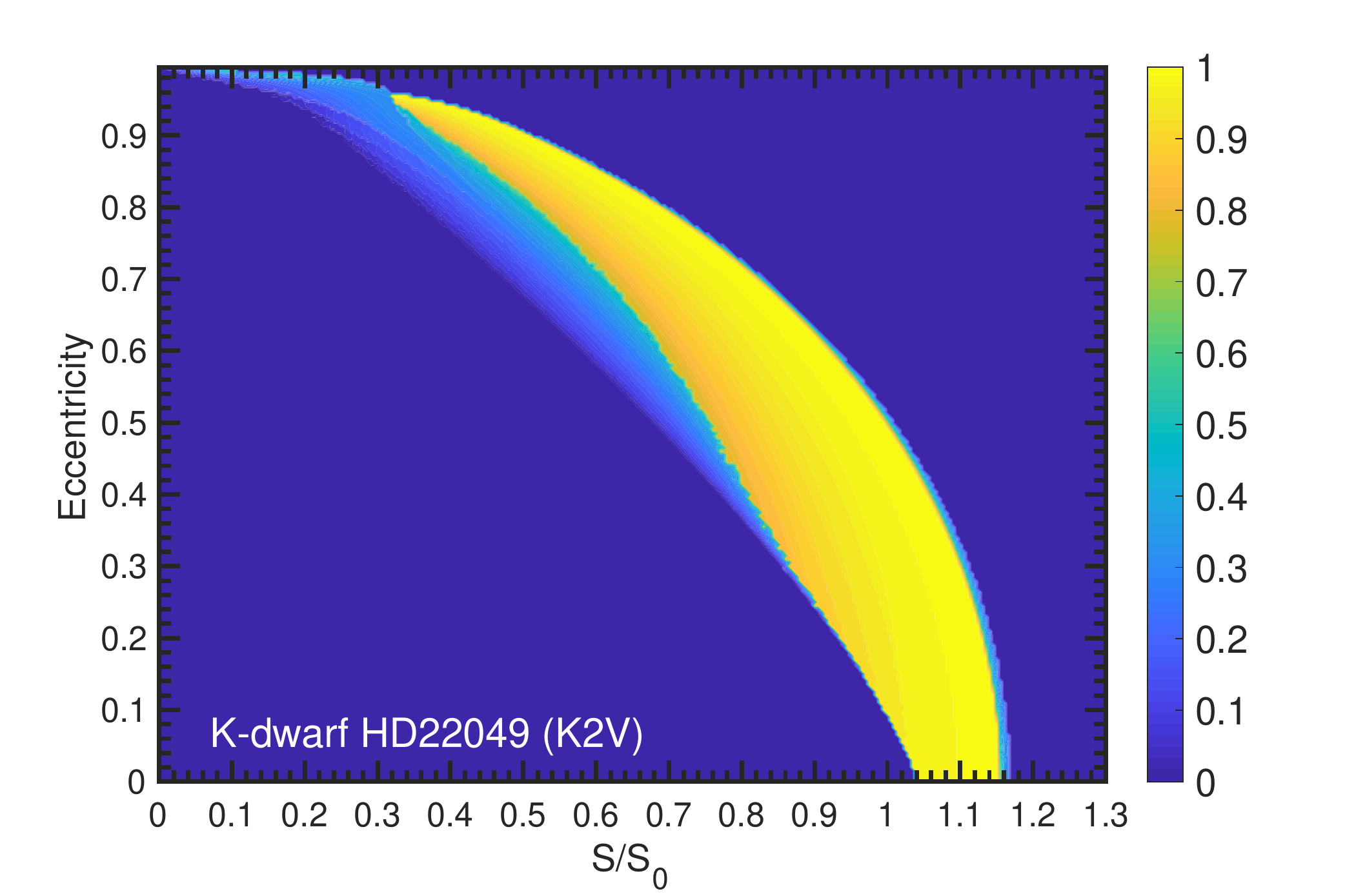}\\
\includegraphics [scale=0.42]{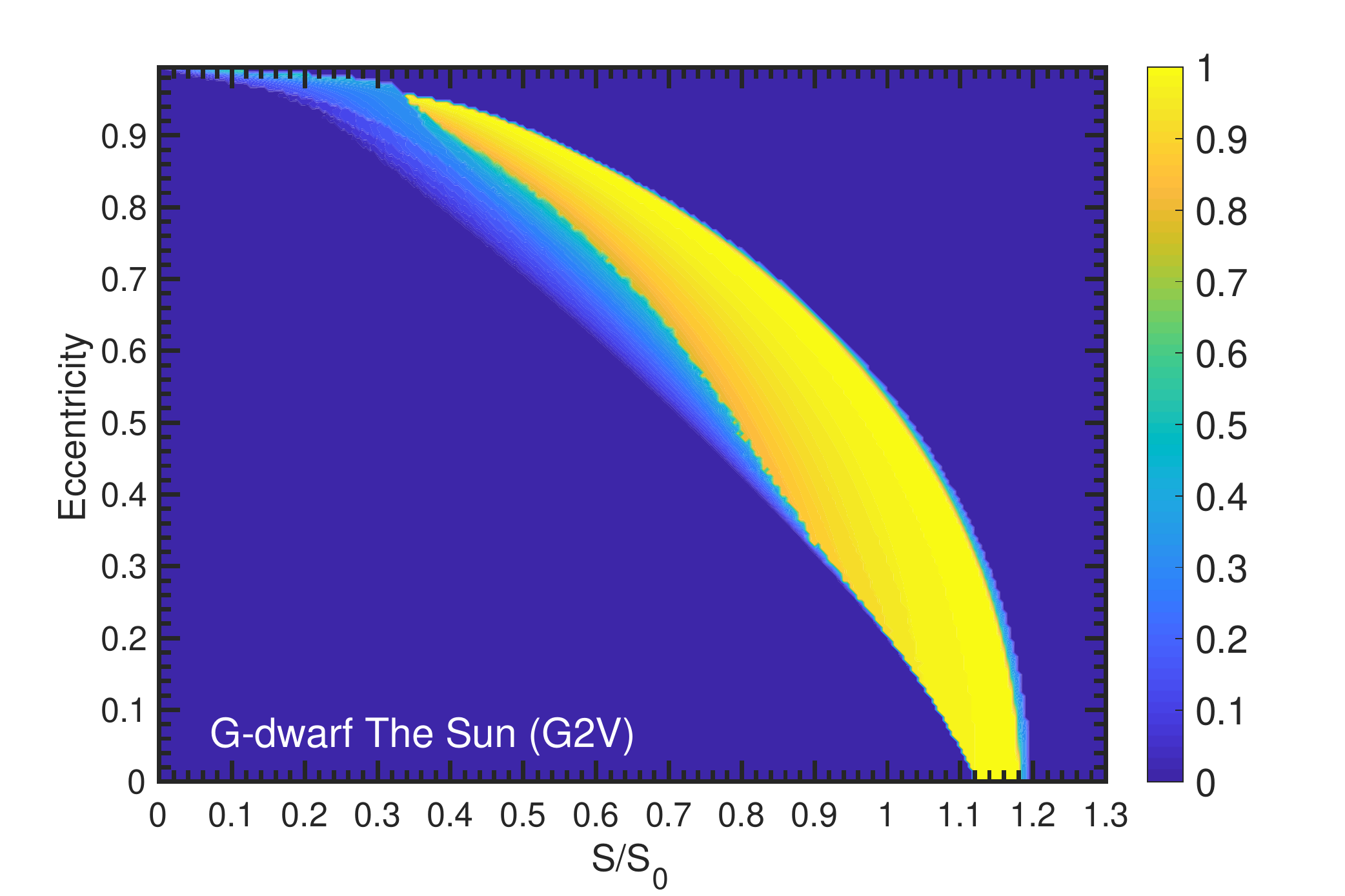}
\includegraphics [scale=0.42]{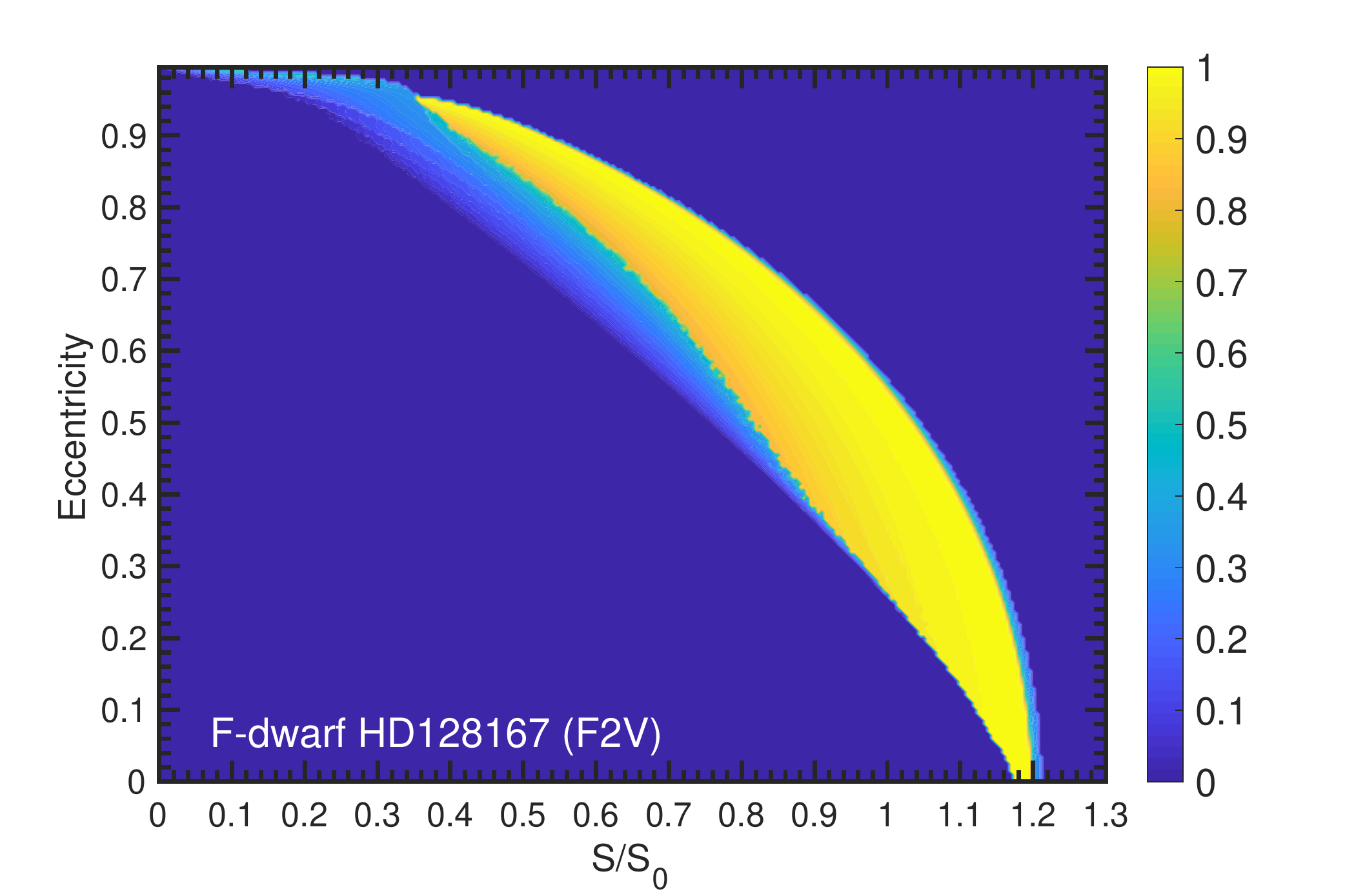}\\
\caption{Fractional Habitability on the M-,K-,G- and F-dwarf planet with varying eccentricity and instellation, after EBM simulations assuming cold start conditions.}
\end{center}
\end{figure}

\begin{figure}
\begin{center}
\includegraphics [scale=0.42]{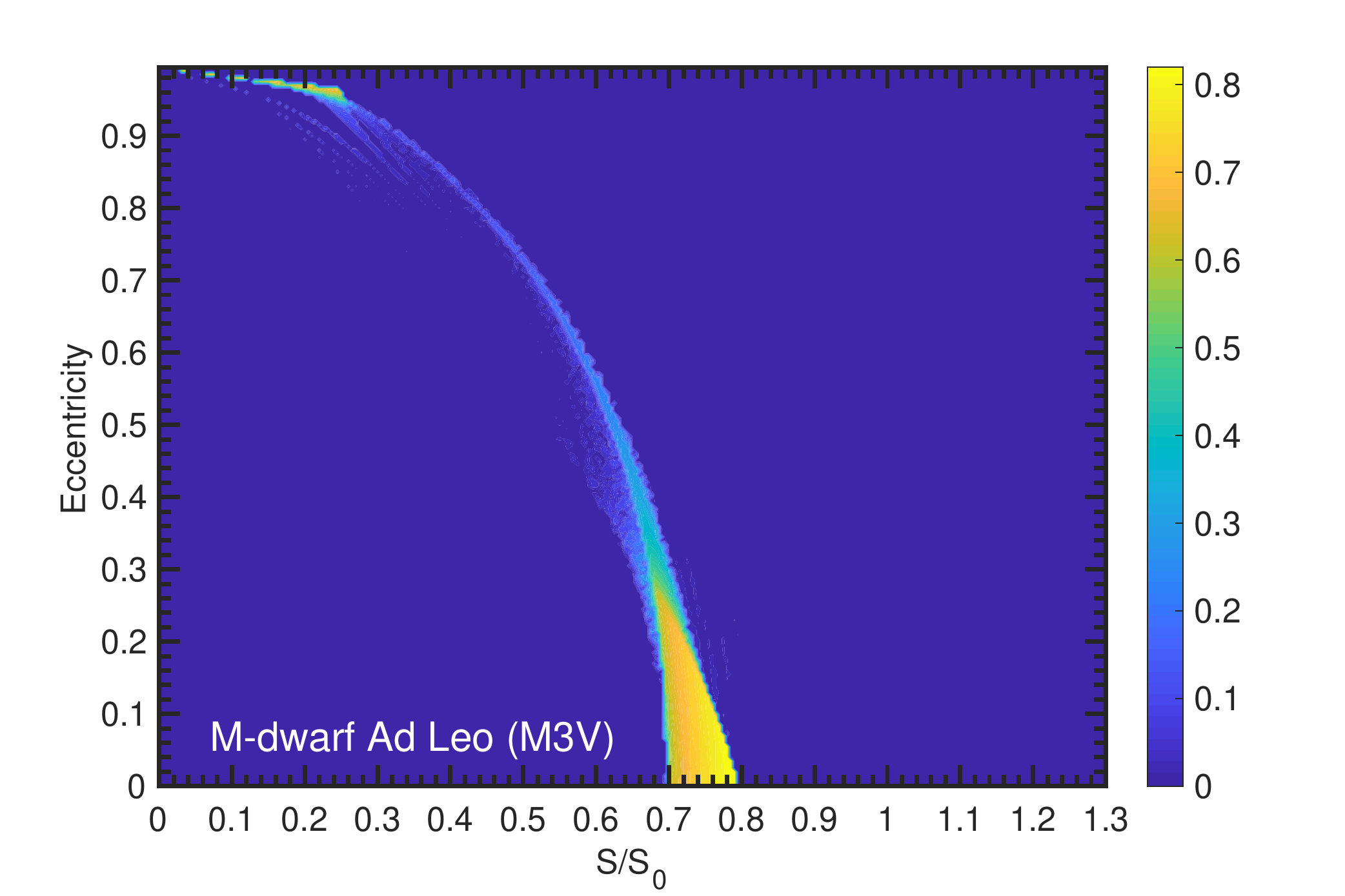}
\includegraphics [scale=0.42]{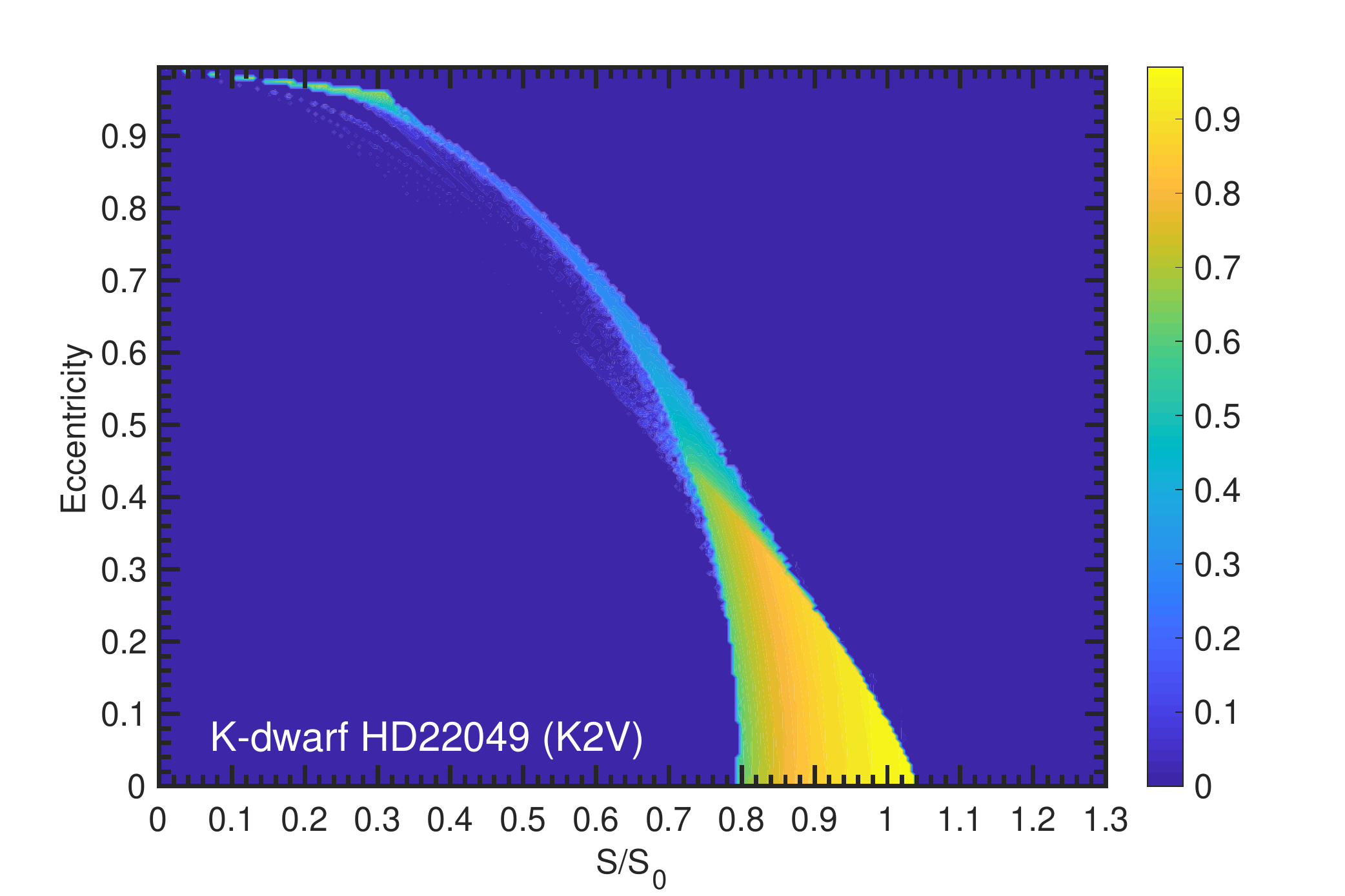}\\
\includegraphics [scale=0.42]{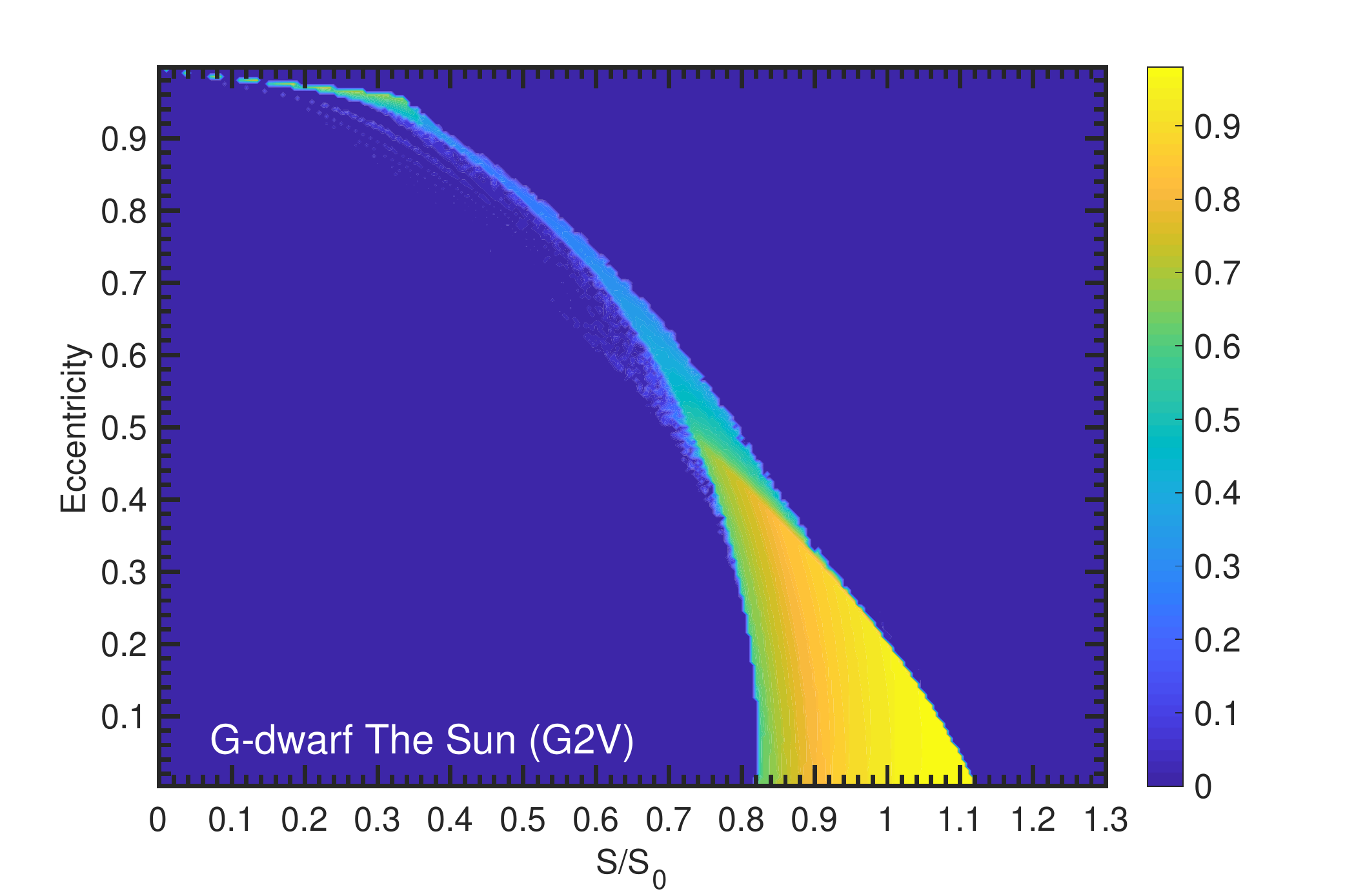}
\includegraphics [scale=0.42]{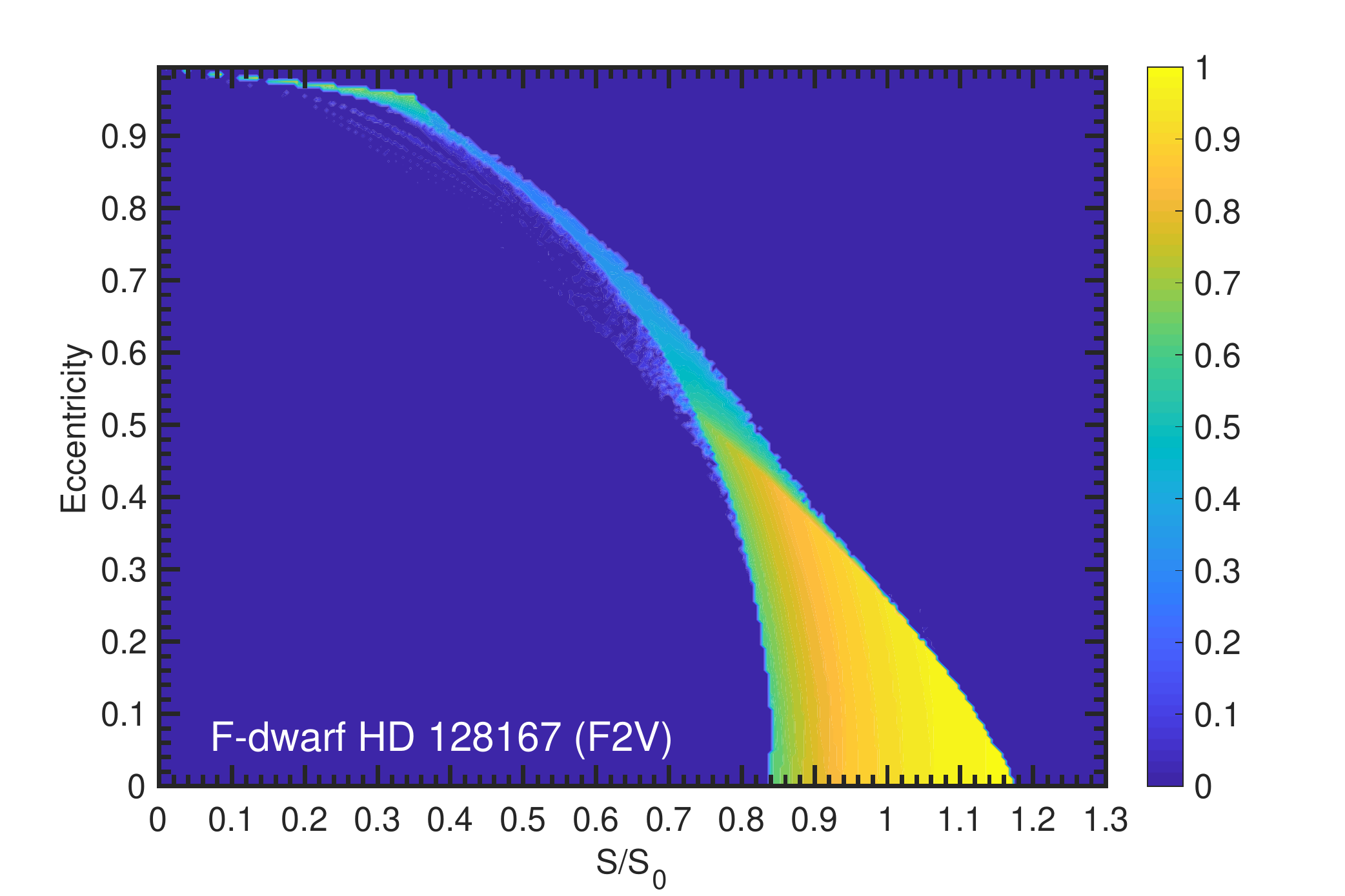}\\
\caption{Warm - cold start difference in the fractional habitability on the M-,K-,G- and F-dwarf planet with varying eccentricity and instellation.}
\end{center}
\end{figure}

\begin{figure}
\begin{center}
\includegraphics [scale=0.5]{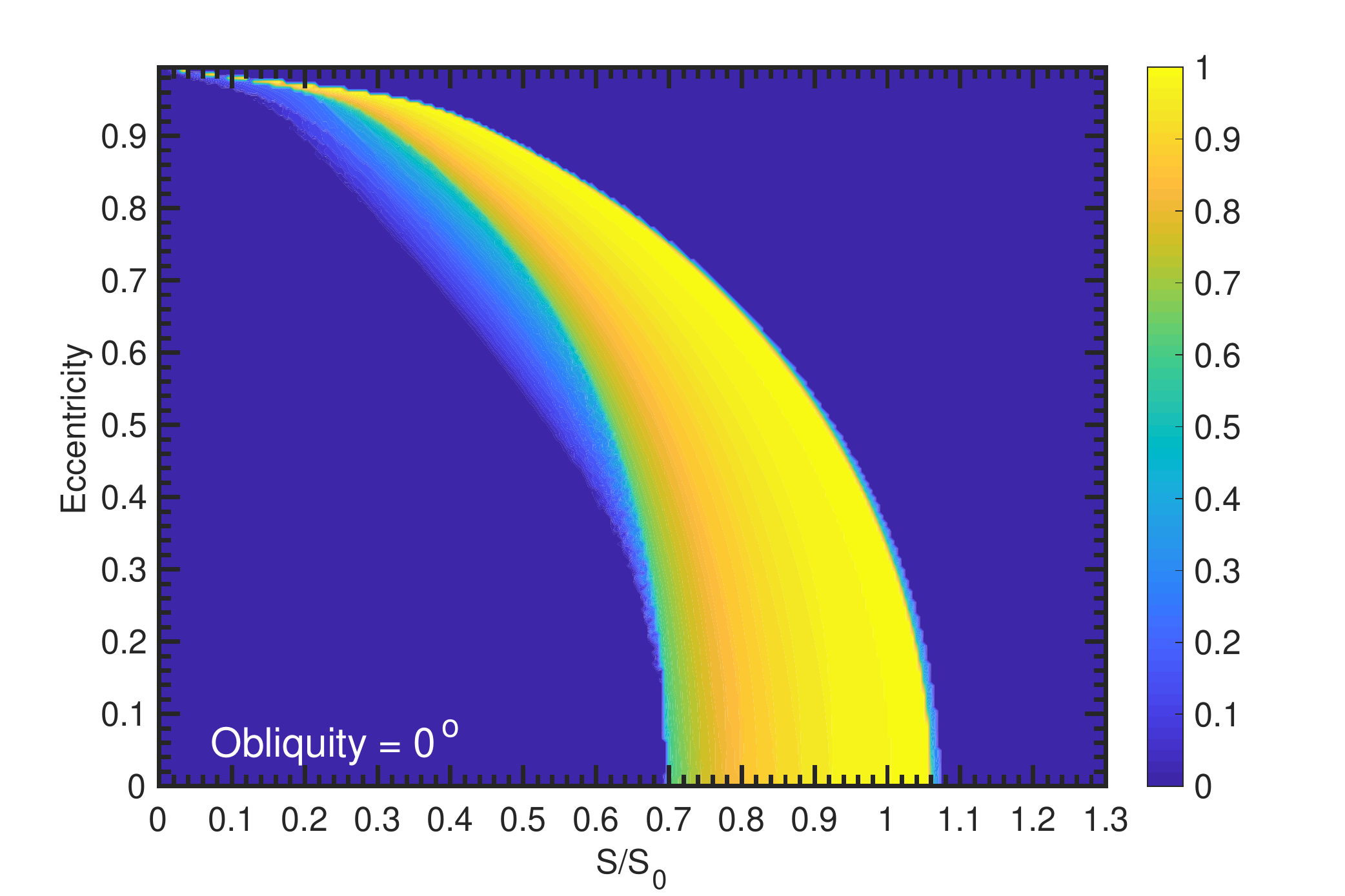}\\
\includegraphics [scale=0.5]{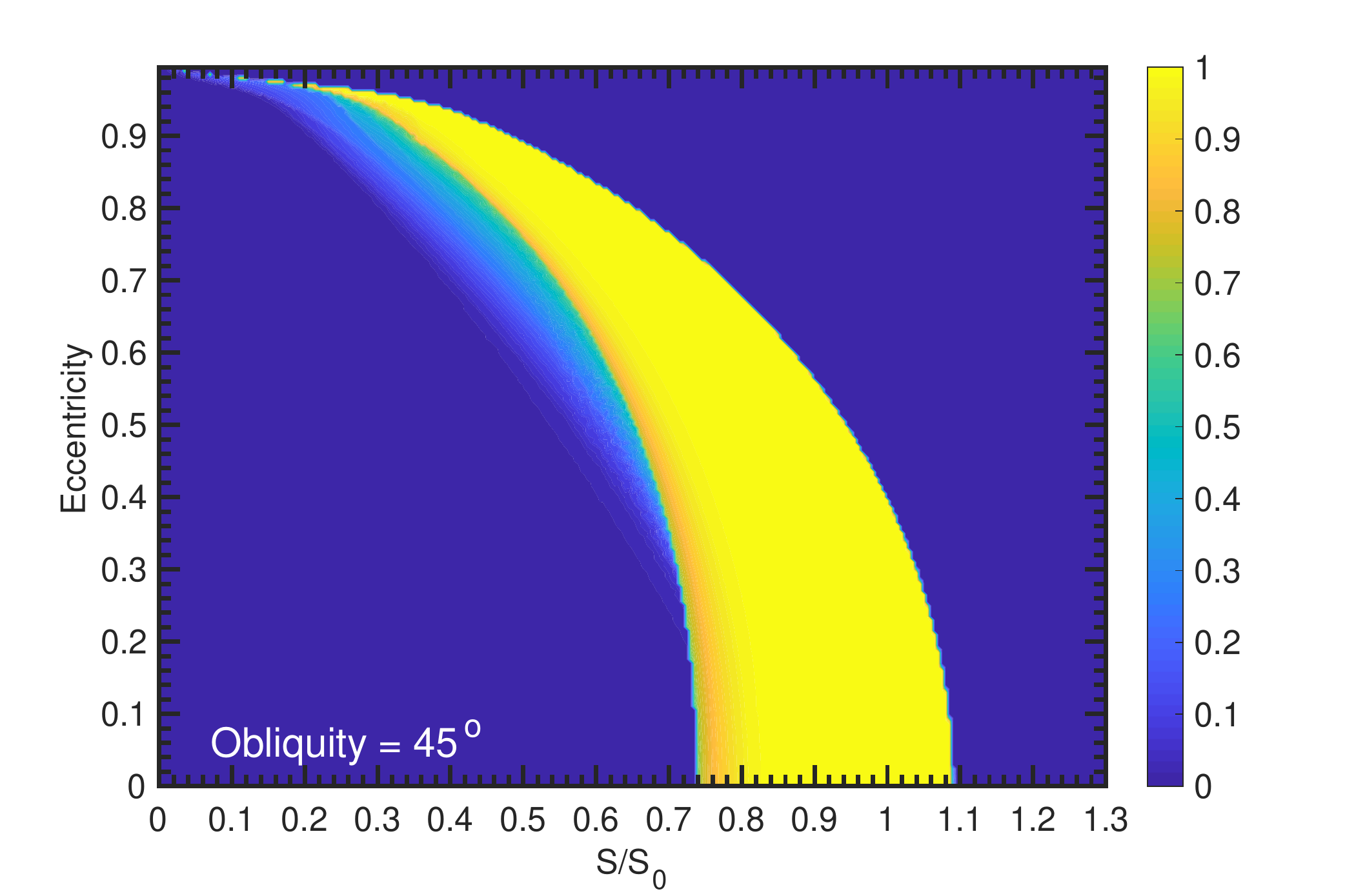}\\
\includegraphics [scale=0.5]{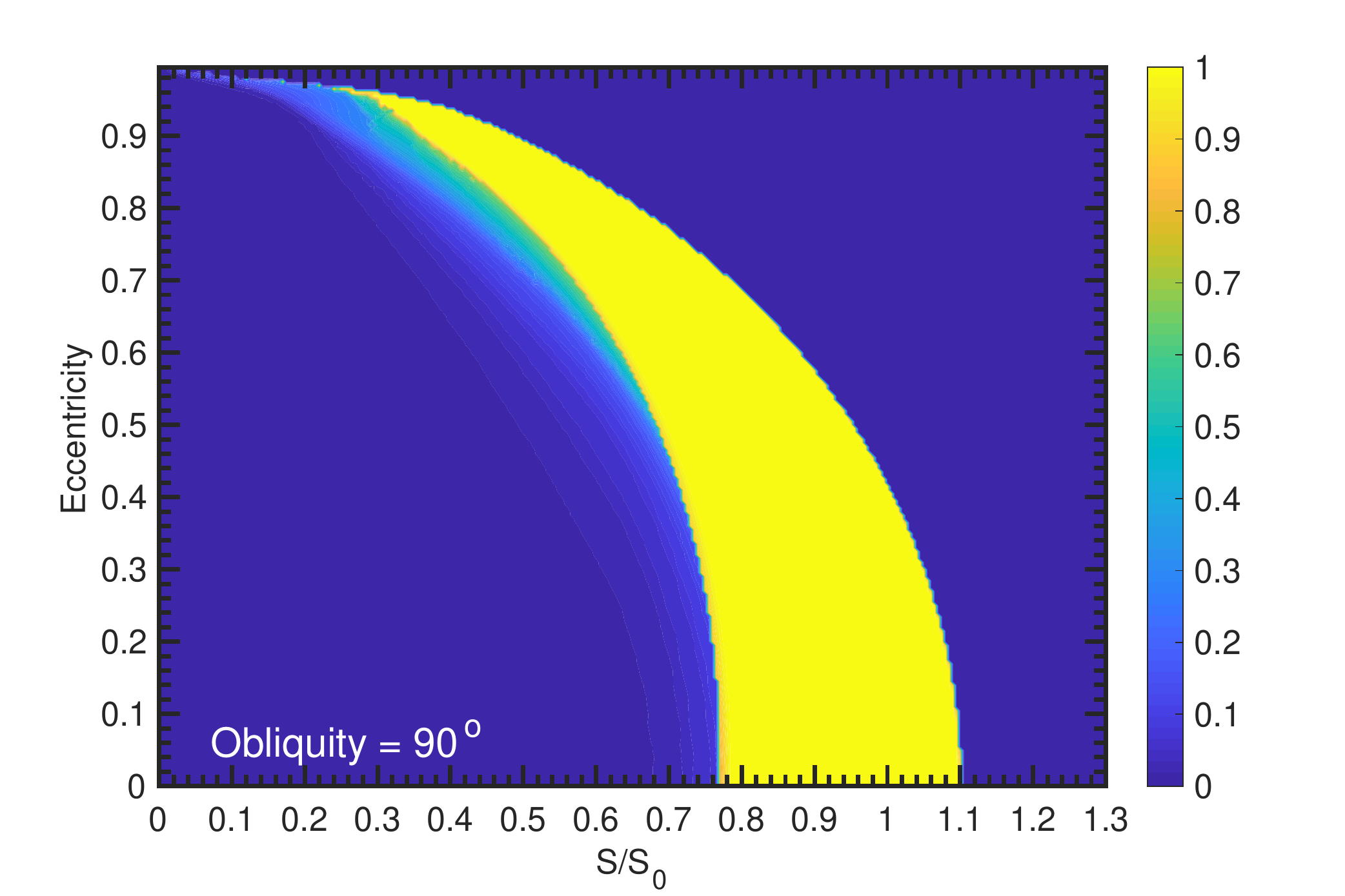}\\
\caption{Fractional Habitability on an M-dwarf planet with $0^{\circ}, 45^{\circ}, 90^{\circ}$ obliquity, after EBM simulations assuming warm start conditions. }
\end{center}
\end{figure}

\end{document}